\documentclass[varenna]{cimento}
\usepackage{epsfig}
\usepackage{graphics}
\usepackage{subfigure}
\def\inseps#1#2{\def\epsfsize##1##2{#2##1} \centerline{\epsfbox{#1}}}
\begin{document}

\begin{center}

{\Large
{\bf Geometrical aspects of protein folding } \\}

\vskip 2.0cm
{\sc Jayanth R. Banavar$^{1}$, Amos Maritan$^{2}$, Cristian
Micheletti$^{2}$ \& Flavio Seno$^{3}$}

\vskip 2.0cm

\end{center}

$^{1}$ Department of Physics and Center for Materials Physics,
104 Davey Laboratory,\\ The Pennsylvania State University, University
Park, Pennsylvania 16802, USA \\

$^{2}$ International School for Advanced Studies (S.I.S.S.A.), \\
Via Beirut 2-4, 34014 Trieste, Istituto Nazionale per la Fisica della
Materia\\ and the Abdus Salam International Center for Theoretical
Physics, Trieste, Italy \\

$^{3}$ 
Istituto Nazionale per la Fisica della Materia (INFM) and\\
 Dipartimento di Fisica G.Galilei, via Marzolo 8, 35131 Padova , Italy \\

\vspace{1.5truecm}

\section{Scope of the lectures}

These lectures will address two questions. Is there a simple
variational principle underlying the existence of secondary motifs in
the native state of proteins? Is there a general approach which can
qualitatively capture the salient features of the folding process and
which may be useful for interpreting and guiding experiments? Here,
we present three different approaches to the first question, which
demonstrate the key role played by the topology of the native state of
proteins. The second question pertaining to the folding dynamics of
proteins remains a challenging problem -- a detailed description
capturing the interactions between amino acids among each other and
with the solvent is a daunting task. We address this issue
building on the lessons learned in tackling the first question and
apply the resulting method to the folding of various proteins
including HIV protease and membrane proteins. The results that will be
presented open a fascinating perspective: the two questions appear to
be intimately related. The variety of results reported here all
provide evidence in favour of the special criteria adopted by nature
in the selection of viable protein folds, ranging from optimal
compactness to maximum dynamical and geometrical accessibility of the
native states.

\noindent {\bf Acknowledgments}. We are indebted to Fabio Cecconi,
Alessandro Laio, Enzo Orlandini, Gianni Settanni and
Antonio Trovato 
who have contributed to the results discussed in these lectures.

\section{Introduction}
\label{sec:intro}

A fascinating and open question challenging biochemistry, physics and
even geometry is the presence of highly regular motifs such as
$\alpha$-helices and $\beta$-sheets in the folded state of biopolymers
and proteins. Stimulating explanations ranging from chemical
propensity to simple geometrical reasoning have been invoked to
rationalize the existence of such secondary structures.

The realization that proteins have secondary structures arose with
early crystallographic studies and the brilliant deduction of Pauling
et al. \cite{Pauling} of the ability of an $\alpha$-helix of the
correct pitch to accomodate hydrogen bonds, thus promoting its
stability. Inspired by the findings of Pauling, helix-coil transition
models have been used to study the thermodynamics of helix formation
\cite{hc}. 
It is interesting to note, however, that
the number of hydrogen bonds is nearly the same when a sequence is in
an unfolded structure in the presence of a polar solvent or in its
native state rich in secondary structure content \cite{Hunt}. It has
also been suggested that the $\alpha$-helix is an energetically
favorable conformation for main-chain atoms but the side-chain suffers
from a loss of entropy \cite{Hunt,Aurora}. Nelson {\em et al.}
\cite{Nelson} have shown both numerically and experimentally that
non-biological oligomers fold reversibly like proteins into a specific
three-dimensional structure with high helical content driven only by
solvophobic interactions.

Recent studies have attempted to explain the emergence of secondary
structure in proteins from geometrical principles rather than invoking
detailed chemistry. Despite the concerted efforts of several groups, a
simple general explanation remains elusive. A very natural line of
investigation was undertaken by Yee {\em et al.} \cite{ChDi}, Hunt
{\em et al.} \cite{Hunt}, and Socci {\em et al.} \cite{On} who
focused on the spontaneous emergence of secondary content from the
mere requirement of overall compactness of homopolymeric chains. Their
findings ruled out any significant relationship between the two, a
fact also corroborated by the recent study of the kinetics of
homopolymer collapse, where no evidence was found for the formation of
local regular structures \cite{halper}. Despite the failure, this
approach is particularly interesting due to the fact that optimal
packing is a fundamental and fascinating problem in contexts ranging
from every day like to atomic physics. Perhaps, the best known packing
problem is the one introduced by Kepler nearly fours centuries ago,
concerning the optimal packing of sphere. However the packing of
independent objects, like spheres, must be treated differently from
the case of objects connected in a chain, such as beads in a string
(an idealization of peptide chains). In Sec. \ref{sec:tube}, the
packing problem is generalized to such chains and, remarkably, if one
requires optimal packing uniformly along the chain, then a particular
type of helix becomes the solution and, furthermore, it has the same
geometrical characteristics as $\alpha$-helices found in proteins.

In addition to packing considerations, dynamical effects also play a
significant role when rapid packing/unpacking is entailed, as in the
formation of amorphous glasses where crystallization is dynamically
thwarted or in the more familiar problem of packing clothes in one's
suitcase. The same question may be asked for protein-like structure.
The fact that they contain motifs which are optimally compact, does
not imply that they can be easily reached from unfolded states. It is,
however, widely believed that native states are, in general, highly
accessible from the kinetic point of view. To investigate this issue
we formulate, in Sec. \ref{sec:fast}, a dynamical variational
principle for selection in conformation space based on the requirement
that the backbone of the native state of biologically viable polymers
be rapidly accessible from the denatured state. The variational
principle is shown to result in the emergence of helical order in
compact structures, revealing a surprising accord with the compactness
requirement discussed above.

Still concerning the folding dynamics, there are two key aspects
distinguishing a protein from a generic heteropolymer: the specially
selected sequence of amino acids and the three-dimensional structure
that it folds reversibly into. Nature uses a rich repertory of twenty
kinds of amino acids with sometimes major and at other times subtle
differences in their interactions with the solvent and with each other
in order to design sequences that fit the putative native state with
minimal frustration \cite{Woly}. The chosen sequences are such that
their target native states are reached through a funnel-like landscape
\cite{9,LMO,funnel,DC} which facilitates the harmonious fitting
together of pieces to form the whole. The three-dimensional structure
impacts on the functionality of the protein and a fascinating issue is
the elucidation of the selection mechanism in conformation space that
picks out certain viable structures from the innumerable ones with a
given compactness. Earlier studies have shown that there is a direct
link between viable native conformations and high designability
\cite{LG,Li}.

A fruitful and general strategy for the study of protein folding would
be to extract information on the folding process directly from the
topology of the native state. This problem will be elucidated in
Sec. \ref{sec:dos} and applied to HIV protease and in
Sec. \ref{sec:mp} to membrane proteins. It will be shown that the
natural folds of proteins have a much larger density of nearby
structures than generic (artificial) conformations of the same
character and that the exceedingly large geometrical accessibility of
natural proteins may be related to the presence of secondary motifs
\cite{19}. It will be shown that a study of the influence of native
state topology on the folding process can reveal information about the
sites that are crucial to the folding process itself. As an
application, we shall identify such sites for three proteins: 2ci2,
barnase and HIV-1 Protease and show that they correlate very well with
the key folding sites identified in experiments.

In Sec. \ref{sec:mp}, a general model based on topological properties
of the native state is introduce to decipher the folding of membrane
proteins.  Nearly a quarter of genomic sequences and almost half of
all receptors that are likely to be targets for drug design
\cite{editorial} are integral membrane proteins.  Understanding the
detailed mechanisms of their foldinging mechanism is a largely
unsolved, key problem in structural biology.  By using our geometrical
approach we can investigate the equilibrium properties and the folding
kinetics of a two helix bundle fragment (comprising $66$
amino-acids) of Bacteriorhodopsin. Once again, the approach seems to be
extremely powerful and it appears to provide an efficient framework
for understanding the variety of folding pathways of transmembrane
proteins.


\section{Optimal shape of a compact polymeric chain}
\label{sec:tube}

A fundamental problem in every day life is that of packing with
examples ranging from fruits in a grocery, clothes and personal
belongings in a suitcase, atoms and colloidal particles in crystals
and glasses, and amino acids in the folded state of proteins
\cite{Sloane,Mackenzie,Wood,Car,Cipra}. The simplest problem in
packing consists of determining the spatial arrangement that
accomodates the highest packing density of its constituent entities
with the result being a crystalline structure.

A classic problem is the determination of the optimal arrangement of
spheres in three dimensions in order to achieve the highest packing
fraction. This problem first posed by Kepler has attracted much
interest culminating in its recent rigorous mathematical solution
\cite{Sloane,Mackenzie} that the answer for infinite systems is a
face-centred-cubic lattice. This simply stated problem has had a
profound impact in many areas \cite{Wood,Car,Cipra}, ranging from the
crystallization and melting of atomic systems, to optimal packing of
objects and subdivision of space.

The close-packed hard sphere problem is simply stated: given $N$ hard
spheres of radius $R$, what is the arrangement which can be enclosed
in the minimum volume, e.g. a cube of side $L$. This is solved by
reformulating the problem, more convenient for numerical
implementation, as the determination of the arrangement of a set of
$N$ points in a cube of linear size, $L_0$, that results in the
minimum of half the distance between any pair of points or between the points and
walls of the container, denoted by $r_{min}$, being as large as
possible \cite{Stewart}. The linear size associated with the region
enclosing the hard spheres follows from dimensional analysis and it is
given by $L = R L_0/r_{min}$, from which it follows that maximizing
$r_{min}$ is equivalent to minimizing $L$. It is notable that the
resulting `bulk' optimal arrangement in the large $N$ limit exhibits
translational invariance in that, far from the boundaries, the local
environment is the same for all points. In dimension $d=2$ and $d=3$
this corresponds to triangular and face-centred-cubic lattices
respectively.

Biopolymers like proteins, DNA and RNA have three dimensional
structures which are rather compact. Furthermore, they are the result
of evolution and one may think that their shape may satisfy some
optimality criterion. This naturally leads one to consider a
generalization of the packing problem of hard spheres to the case of
flexible tubes with a uniform cross section. The packing problem then
consists in finding the tube configuration which can be enclosed in
the minimum volume without violating any steric contraints.

The problem can alternatively be formulated in a very simple and elegant way
in terms of the curve which is the axis of the tube (the analog of
the sphere centers in the hard sphere packing problem) \cite{tubone}.
Consider a string (an open curve) in three dimensions. We will utilize
a geometric measure \cite{Buck1} of the curve, the `rope-length',
defined as the arc length measured in units of the thickness, which
has proved to be valuable in applications of knot theory
\cite{Buck1,Knots1,Knots2,Thick3,Buck2,Cantarella}. The thickness
$\Delta$ denotes the maximum radius of a uniform tube with the string
passing through its axis, beyond which the tube either ceases to be
smooth, owing to tight local bends, or it self-intersects. Our focus
is on finding the optimal shape of a curve of fixed arc length,
subject to constraints of compactness, which would maximize its
thickness, or equivalently minimize its rope length. 

Following the approach of Gonzalez and Maddocks \cite{Thick3}, who
studied knotted strings, we define a global radius of curvature as
follows. The global radius of curvature of the curve at a given point
is computed as the minimum radius of the circles going through that
point and all other pairs of points of the string. It generalizes the
concept of the local radius of curvature (the radius of the circle
which locally best approximates the curve) by taking into account
both local (bending of the string) and non-local (proximity to another
part of the string) effects. For discretized curves, the local radius
of curvature at a point is simply the radius of the circle going
through the point and its two adjoining points. The minimum of all the
global radii then defines the thickness, i.e. the minimum radius of
the circles going through any triplet of discrete points. This
coincides with the previous definition in the continuum limit,
obtained on increasing the number of discretized points (assumed to be
equally spaced) on the curve keeping the string length fixed
\cite{Thick3}. Given a string configuration, the thickness is just
the maximum allowed radius for the cross section of a uniform tube
that has the given curve as its axis \cite{Thick3}.

\begin{figure}
\centering
\subfigure[]{\includegraphics[width=1.2in]{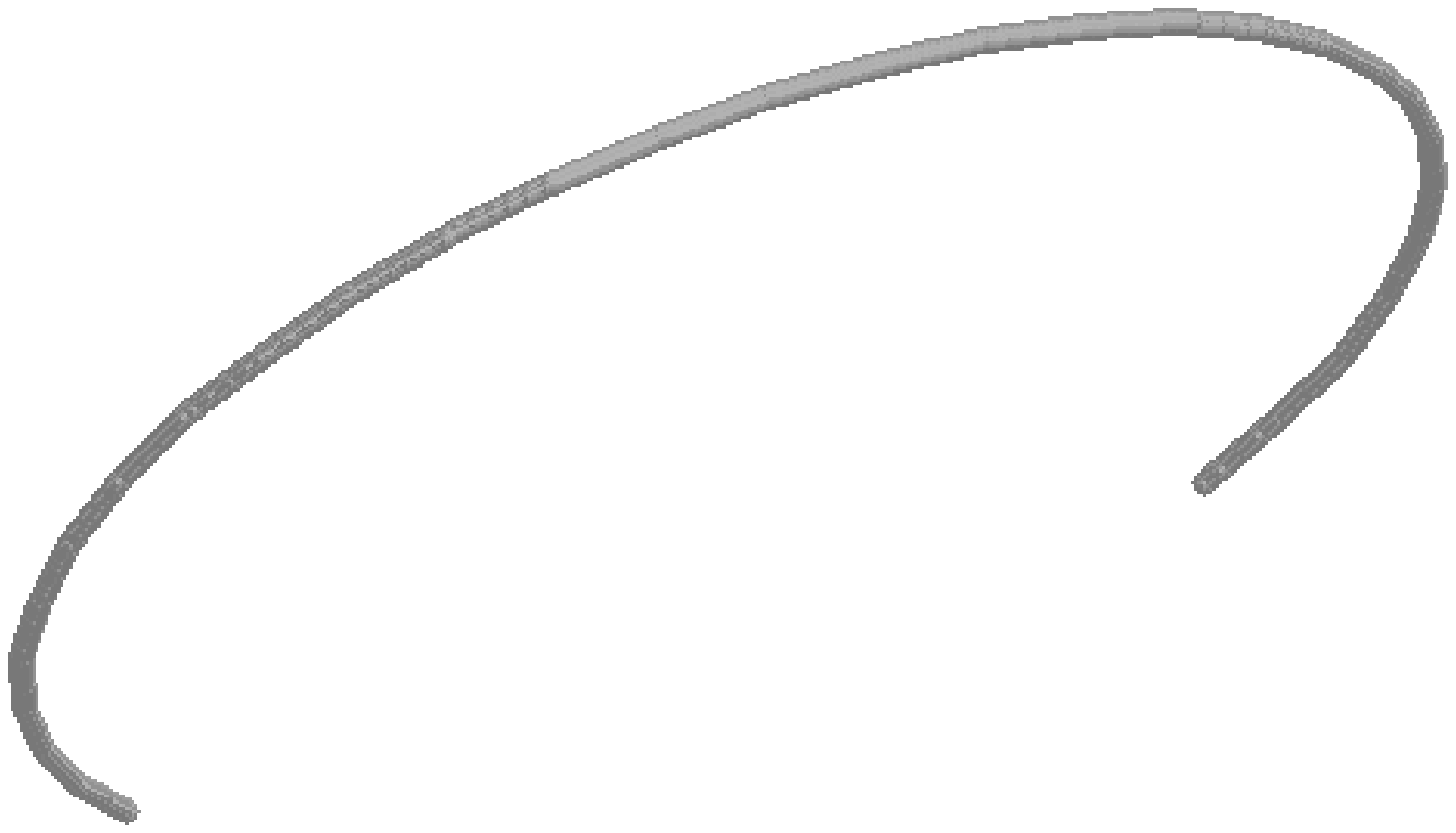}}
\hfill
\subfigure[]{\includegraphics[width=1.2in]{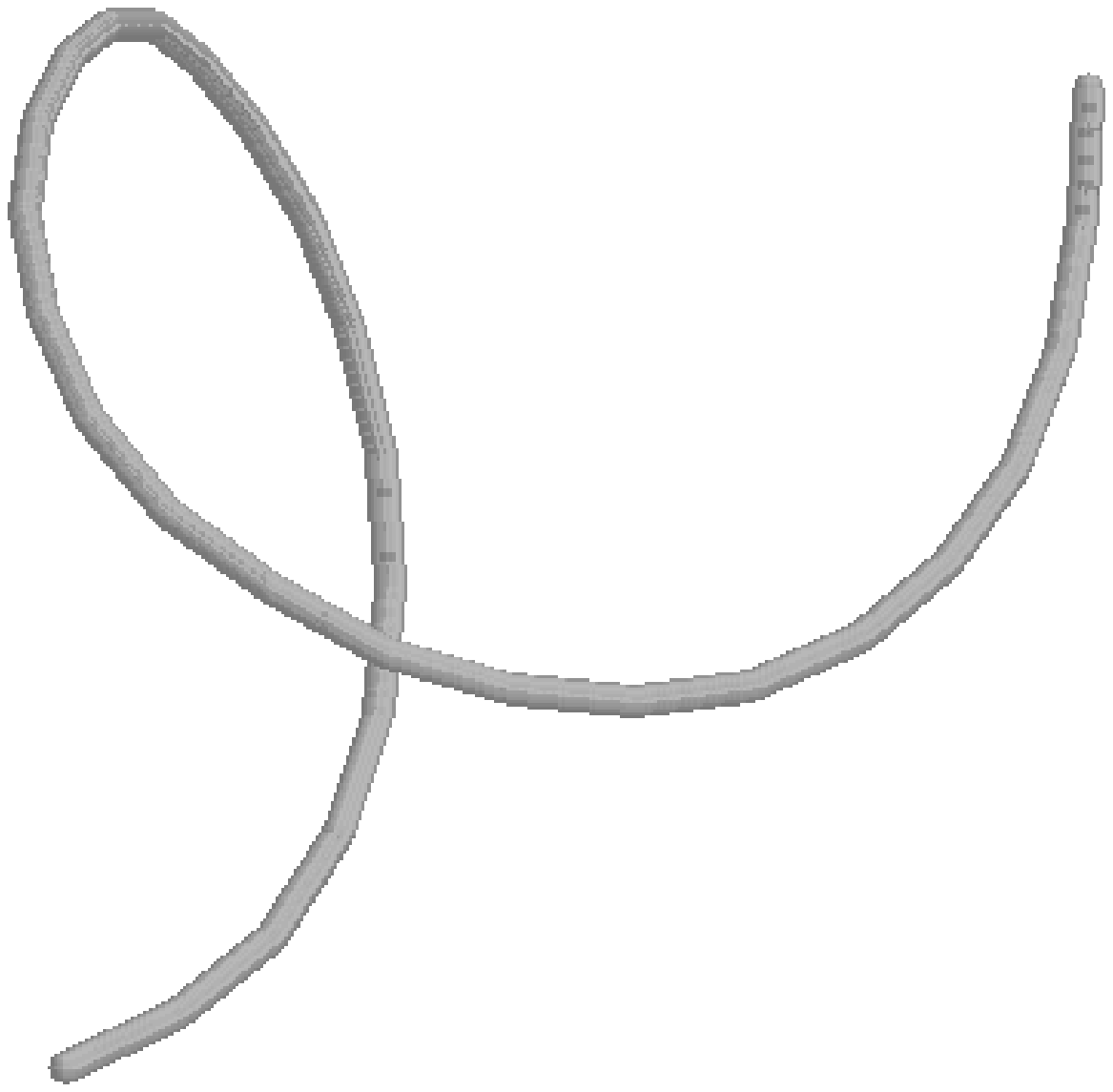}}
\hfill
\subfigure[]{\includegraphics[width=1.2in]{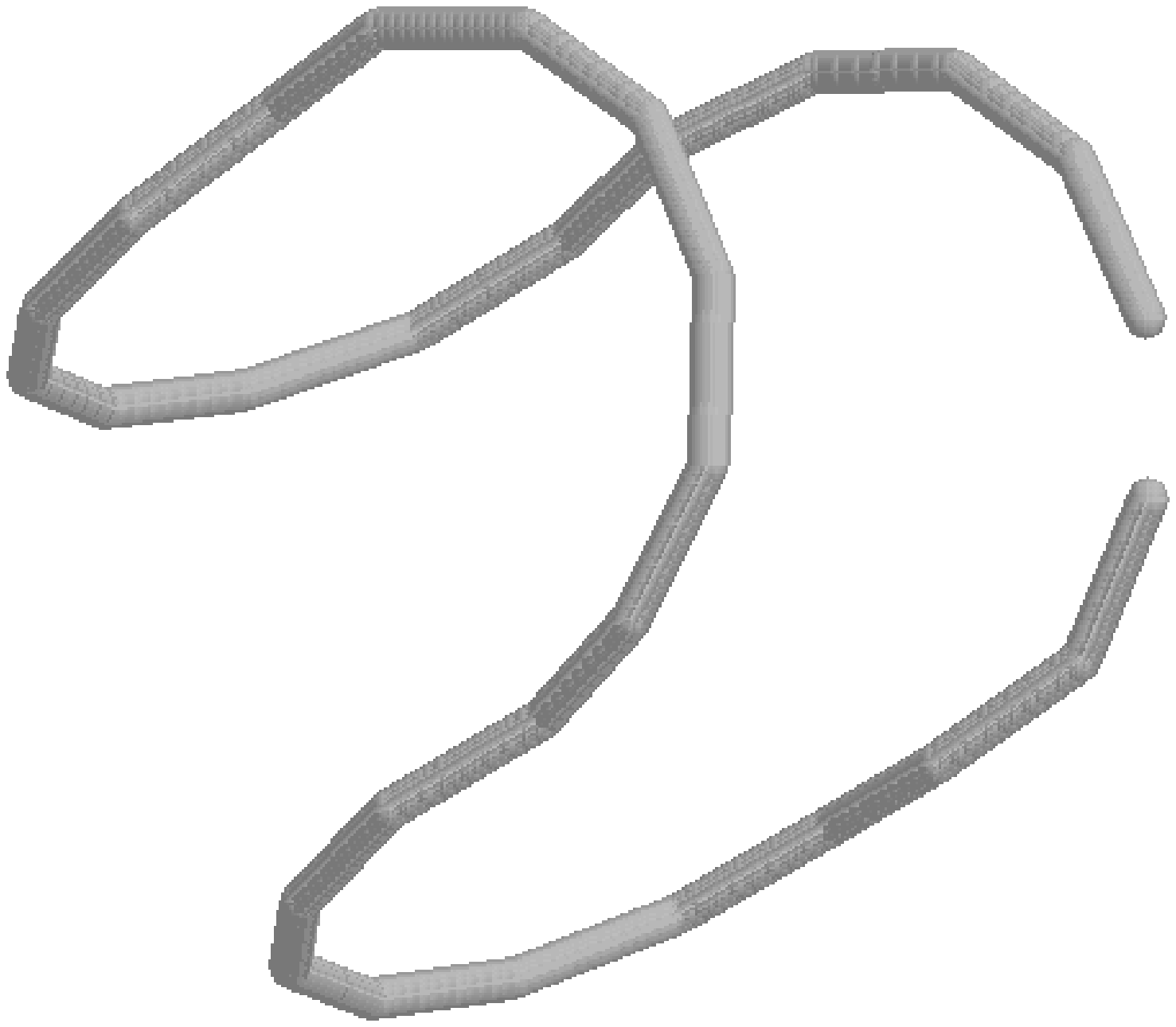}}
\\
\subfigure[]{\includegraphics[width=1.2in]{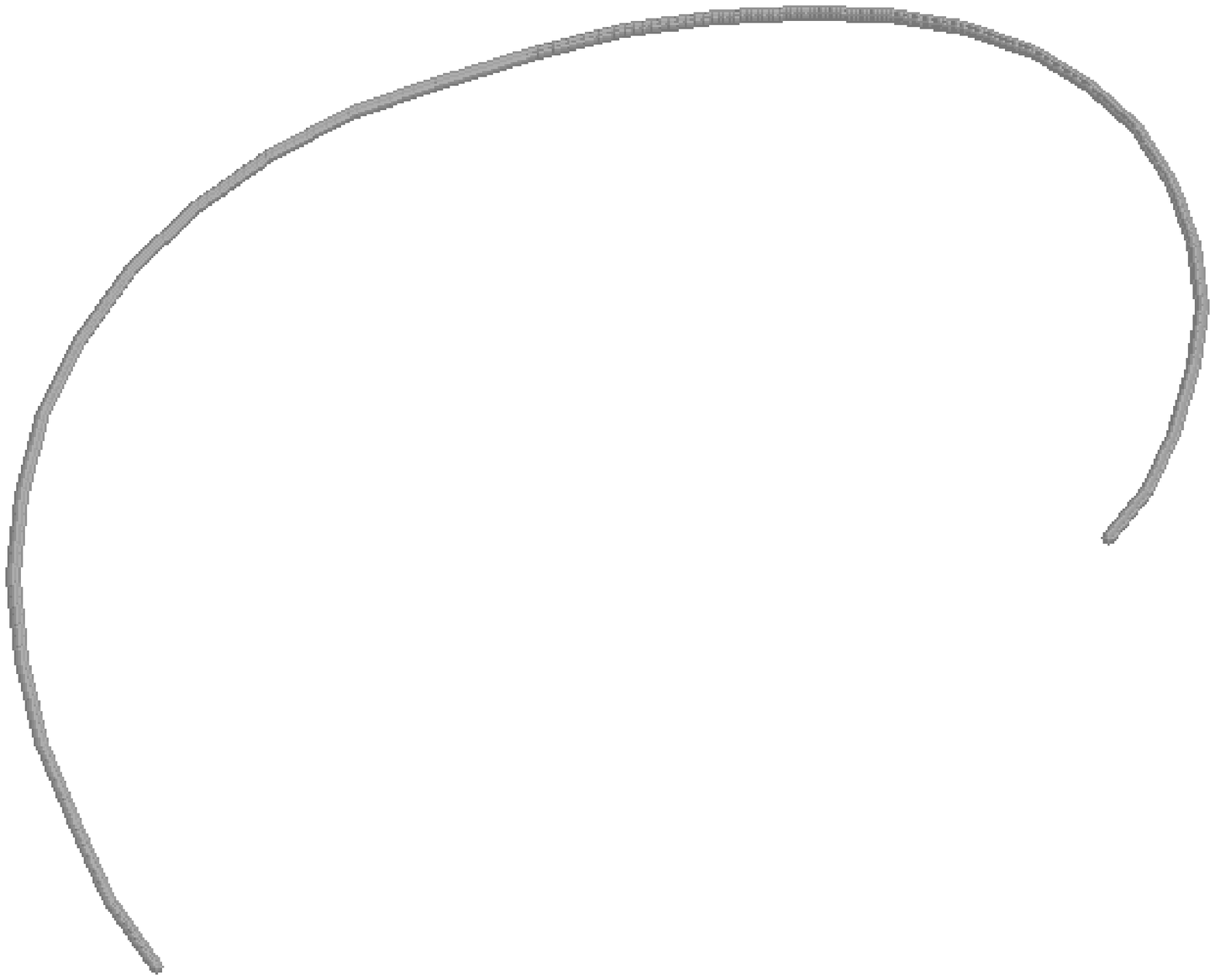}}
\hfill
\subfigure[]{\includegraphics[width=1.2in]{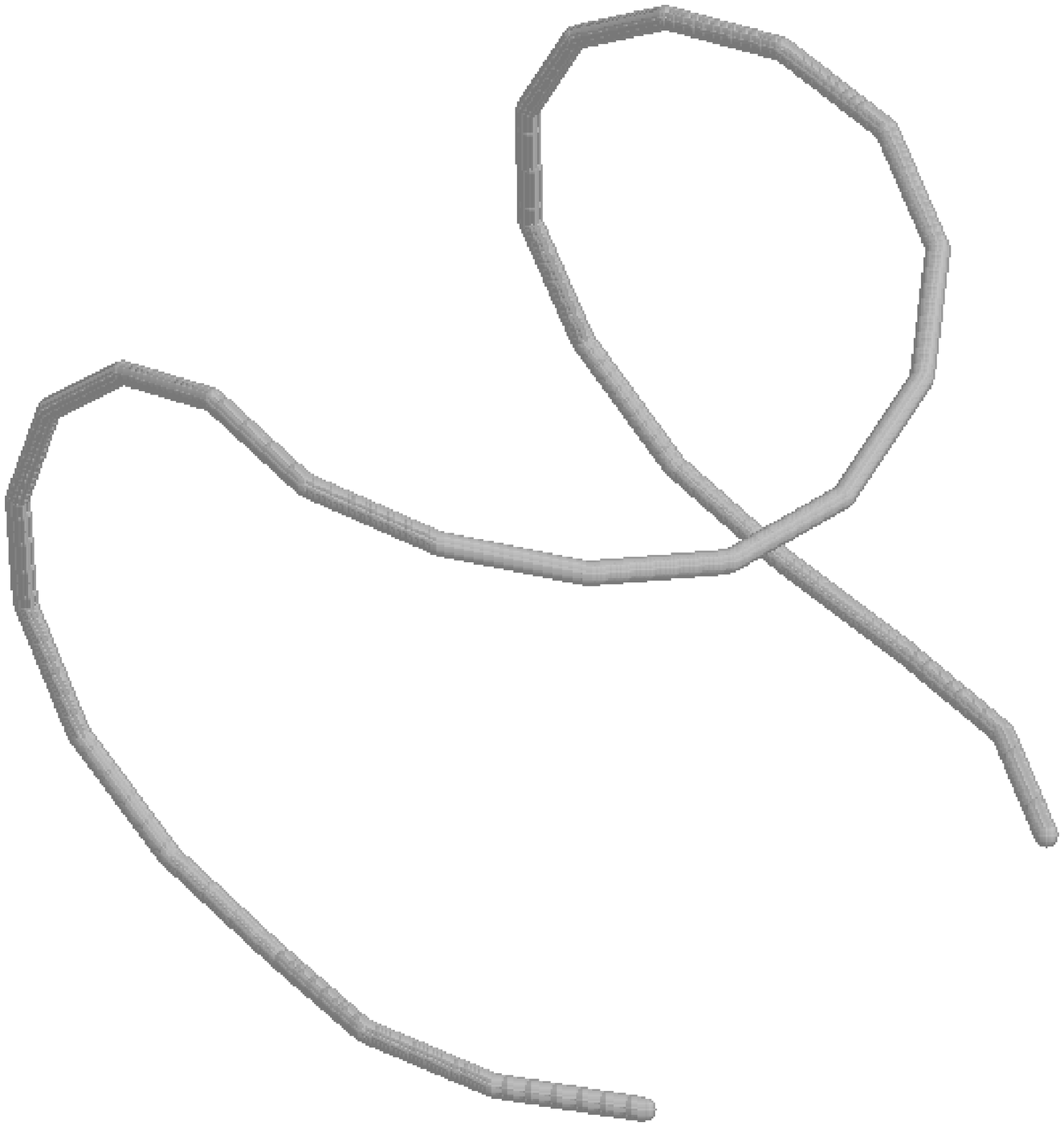}}
\hfill
\subfigure[]{\includegraphics[width=1.2in]{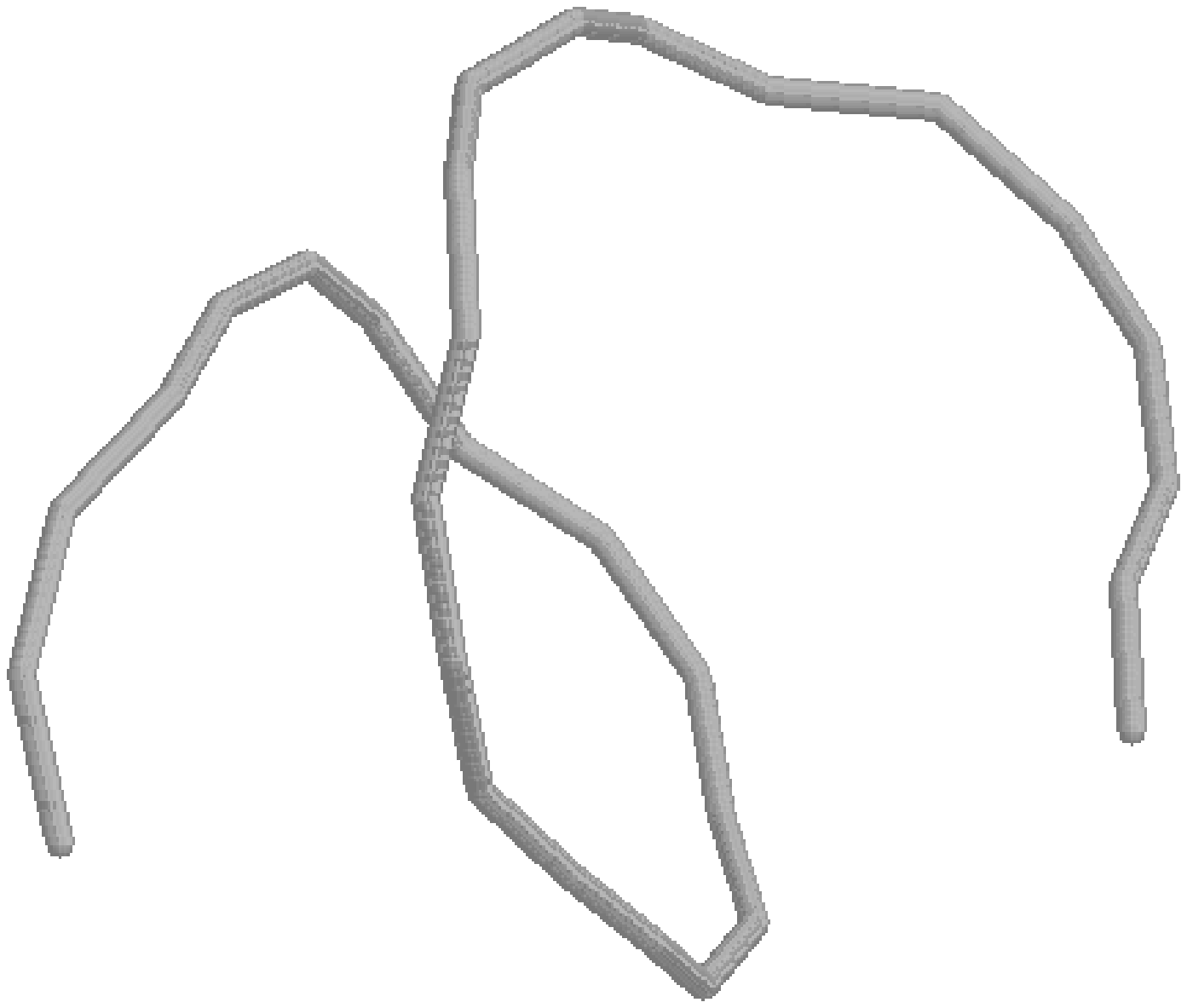}}
\caption{Examples of optimal strings. The strings in
the figure were obtained starting from a random conformation of a
chain made up of $N$ equally spaced points (the spacing between
neighboring points is defined to be 1 unit) and successively
distorting the chain with pivot, crankshaft and slithering moves
commonly used in stochastic chain dynamics \cite{Sokal}. A Metropolis
Monte Carlo procedure is employed with a thermal weight,
$e^{+\Delta/T}$ , where $\Delta$ is the thickness and $T$ is a
fictitious temperature set initially to a high value such that the
acceptance rate is close to 1 and then decreased gradually to zero in
several thousand steps. Self-avoidance of the optimal string
is a natural consequence of the maximization of the thickness.
The introduction of a hard-core repulsion between beads was found
to significantly speed up convergence to the optimal solution and
avoid trapping in self-intersecting structures. We have verified
that the same values (within 1 percent) of the final thickness of the
optimal strings are obtained starting from unrelated initial
conformations. Top row: optimal shapes obtained by constraining
strings of 30 points with a radius of gyration less than $R$. (a) $R =
6.0$, $\Delta = 6.42$ (b) $R = 4.5$, $\Delta = 3.82$ (c) $R = 3.0$,
$\Delta = 1.93$. Bottom row: optimal shapes obtained by confining a
string of 30 points within a cube of side $L$. (d) $L = 22.0$, $\Delta
= 6.11$ (e) $L = 9.5$, $\Delta = 2.3$ (f) $L = 8.1$, $\Delta = 1.75$.}
\label{fig:tubefig1}
\end{figure}

We used several different boundary conditions to enforce the
confinement of the string. The simplest ones discussed here are the
confinement of a curve of length $l$ within a cube of side $L$ or
constraining it to have a radius of gyration (which is the
root-mean-square distance of the discretized points from their centre
of mass) that is less than a pre-assigned value $R$. Even though
different boundary conditions influence the optimal string shape, the
overall features are found to be robust. Examples of optimal shapes,
obtained from numerical simulations, for different ratios of $l/L$ and
$l/R$ are shown in Fig. \ref{fig:tubefig1}. In both cases, two distinct
families of curves, helices and saddles, appear. The two families are
close competitors for optimality and different boundary conditions may
stabilize one over the other. For example, if optimal strings of
fixed length are constrained to have a radius of gyration less than
$R$, then upon decreasing $R$, the curve goes from a regime where the
trivial linear string is curled into an arc, then into a portion of
helix and finally into a saddle. When the string is constrained to lie
within a cube of size $L$, as $L$ decreases first saddles are observed
and then helices.

\begin{figure}
\centerline{\psfig{figure=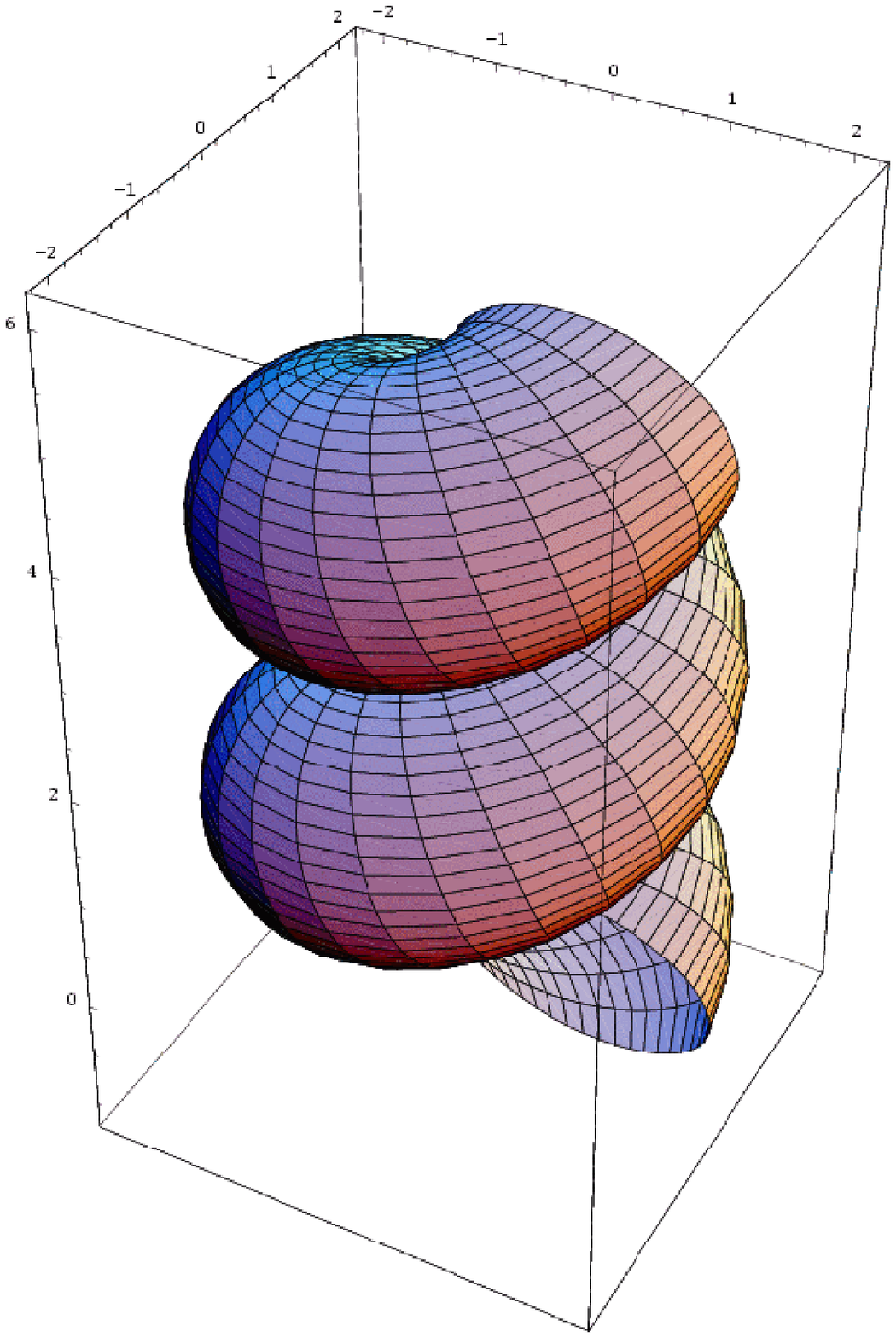,height=5.5in}}
\caption{Shape of the optimal helix. The ratio of the pitch to radius
of the centerline is 2.512.}
\label{fig:tubefig2}
\end{figure}

We have also been able to find bulk-like solutions which are not
influenced by boundary effects. Such solutions can be obtained by
imposing uniform local constraints along the curve. On imposing a
minimum local density on successive segments of the string (for
example, constraining each set of six consecutive beads to have a
radius of gyration that is less than a preassigned value $R$), we
obtained perfectly helical strings, corresponding to discretised
approximations to the continuous helix represented in
Fig. \ref{fig:tubefig2} confirming that this is the optimal
arrangement. Note that, in close analogy with the sphere-packing
problem, a helix has translational invariance along the chain.

In all cases, the geometry of the chosen helix is such that there is
an equality of the local radius of curvature (determined by the local
bending of the curve) and the radius associated with a suitable
triplet of non-consecutive points lying in two successive turns of the
helix. This is a feature that is observed only for a special ratio
$c^{*}$ of the pitch, $p$, and the radius, $r$, of the circle
projected by the helix on a plane perpendicular to its axis. When
$p/r>c^{*}=2.512$ the local radius of curvature, given by $\rho= r
(1+p^2/(2\pi r)^2)$, is lower than the half of the
distance of closest approach of
points on successive turns of the helix. The latter is given by the
first minimum of $1/2 \sqrt{2 - 2 \cos (2 \pi t) + p^2 t^2}$ for
 $t > 0$. Thus $\Delta = \rho$ in this case.

If $p/r<c^{*}$, the global radius of curvature is strictly lower than
the local radius, and the helix thickness is determined basically by
the distance between two consecutive helix turns: $\Delta\simeq p/2$
if $p/r\ll1$. Optimal packing selects the very special helices
corresponding to the transition between the two regimes described
above. A visual example is provided by the optimal helix of
Fig. \ref{fig:tubefig2}.

For discrete curves, the critical ratio $p/r$ depends on the
discretization level. A more robust quantity is the ratio $f$ of the
minimum radius of the circles going through a given point and any two
non-adjacent points and the local radius. For discretized strings,
$f=1$ just at the transition described above, whereas $f>1$ in the
`local' regime and $f<1$ in the `non-local' regime. In our
computer-generated optimal strings, $f$ differed from unity by less
than a part in a thousand.

It is interesting to note that, in nature, there are many instances of
the appearance of helices. For example, many biopolymers such, as
proteins and enzymes, have backbones which frequently form helical
motifs. (Rose and Seltzer \cite{Rose} have used the local radii of
curvature of the backbone as input in an algorithm for finding the
peptide chain turns in a globular protein.) It has been shown
\cite{19} that the emergence of such motifs in proteins (unlike in
random heteropolymers which, in the melt, have structures conforming
to Gaussian statistics) is the result of the evolutionary pressure
exerted by nature in the selection of native state structures that are
able to house sequences of amino acids which fold reproducibly and
rapidly \cite{M} and are characterized by a high degree of
thermodynamic stability \cite{SSK}. Furthermore, because of the
interaction of the amino acids with the solvent, globular proteins
attain compact shapes in their folded states.

\begin{figure}
\centerline{\psfig{figure=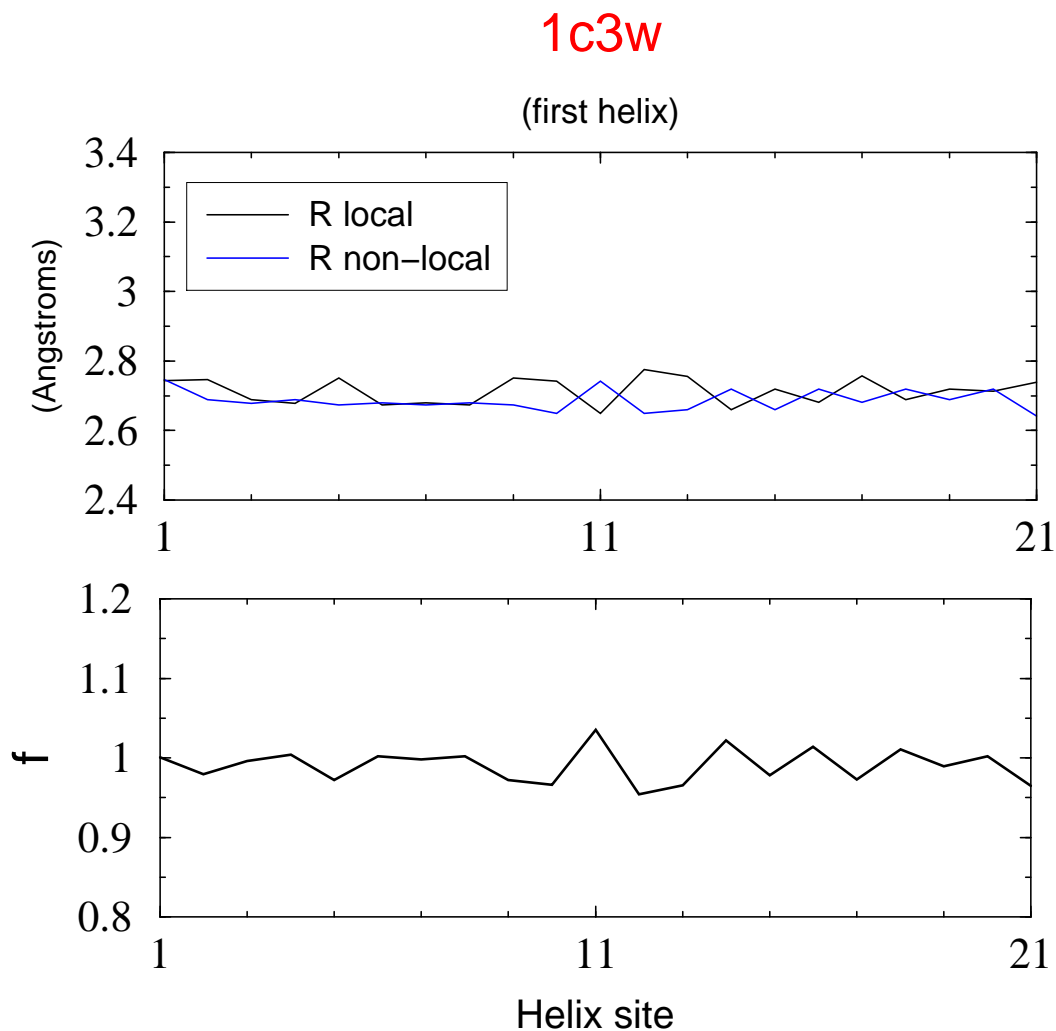,width=5.5in,height=5.5in}}
\caption{Top. Local and non-local radii of curvature for sites in the
first helix of bacteriorhodopsin (pdb code 1c3w). Bottom. Plot of $f$
values for the same sites.}
\label{fig:tubenew}
\end{figure}

It is then natural to measure the shape of these helices and assess if
they are optimal in the sense described here. The measure of $f$ in
$\alpha$-helices found in naturally-occurring proteins yields an
average value for $f$ of $1.03 \pm 0.01$, hinting that, despite the
complex atomic chemistry associated with the hydrogen bond and the
covalent bonds along the backbone, helices in proteins satisfy optimal
packing constraints. An example is provided in Fig. \ref{fig:tubenew}
where we report the value of $f$ for a particularly long
$\alpha$-helix encountered in a heavily-investigated membrane protein,
bacteriorhodopsin.

\begin{figure}
\centerline{\psfig{figure=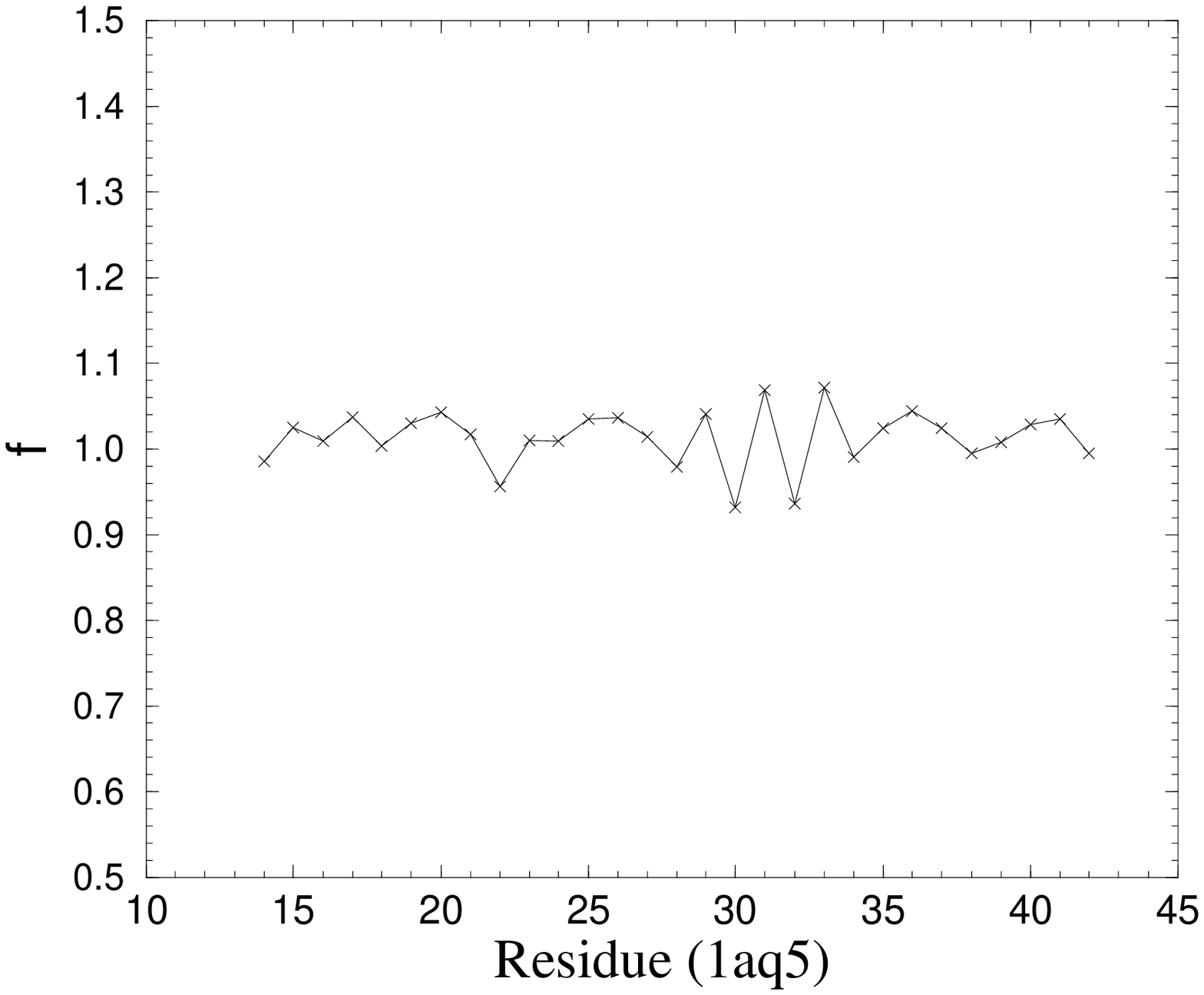,width=5.5in,height=5.5in}}
\caption{Packing of collagen helices. Plot of $f$ values as a
function of sequence position for a single collagen helix (only
$C_\alpha$ coordinates were used to identify the protein
backbone). The same plot for each of the three collagen chains would
simply superimpose. We considered the residues 14-42 from the
structure 1aq5 in the Protein Data Bank.}
\label{fig:tubefig3}
\end{figure}

This result implies that the backbone sites in protein helices have an
associated free volume distributed more uniformly than in any other
conformation with the same density. This is consistent with the
observation \cite{19} that secondary structures in natural proteins
have a much larger configurational entropy than other compact
conformations. This uniformity in the free volume distribution seems
to be an essential feature because the requirement of a maximum
packing of backbone sites by itself does not lead to secondary
structure formation \cite{ChDi,On}. Furthermore, the same result also
holds for the helices appearing in the collagen native state
structure, which have a rather different geometry (in terms of local
turn angles, residues per turn and pitch \cite{Creighton}) from
average $\alpha$-helices. In spite of these differences, we again
obtained an average $f = 1.01 \pm 0.03$ (Fig. \ref{fig:tubefig3}),
very close to the optimal situation.

\vskip 2.0cm
\section{Fast folding polymers and role of secondary motifs}
\label{sec:fast}

Fast packing has been recognized as a central issue for biopolymers,
such as proteins, since the early work of Levinthal \cite{Levin}. 
The packing of strings as defined above and the slope of the optimal
string
we have found should also be kinetically very accessible.
We
postulate a direct connection between the dynamics of rapid folding
and the emergence of secondary motifs in the native state
conformations \cite{M}. In fact, an intuitive approach to
rapid and reproducible folding might be to create neat patterns of
lower dimensional manifolds than the physical space and bend and curl
them into the final folded state. For proteins, secondary structures
such as $\alpha$-helices and $\beta$-sheets are indeed patterns in low
dimensions.

The selection mechanism in structure space is formulated as a
variational principle postulating that, {\em among all possible native
conformations, a protein backbone will attain only those which are
optimal under the action of evolutionary pressure favouring rapid
folding}. Our goal is to elucidate the role played by the bare native
backbone independent of the selection in sequence space and hence of
the (imperfectly-known) inter-amino-acid potentials. We therefore
choose to employ a Go-like model \cite{Go} with no other interaction
that promotes or disfavours secondary structures. The model is a
sequence-independent limiting case of minimal frustration\cite{Woly}
which, for a given target native state conformation, favours the
formation of native contacts -- the energy of a sequence in a
conformation is simply obtained as the negative of the number of
contacts in common with the target conformation. We will work in the
$C_{\alpha}$ representation and consider two non-consecutive amino
acids to be in contact if their separation is below a cutoff $r_0 =
6.5$ \AA (the results are qualitatively similar when slightly
different values of $r_0$ in the range $6-8$ \AA are chosen).

The energy of structure $\Gamma$ in the Go model is given by
\begin{equation}
H(\Gamma) = - {1 \over 2} \sum_{i, j} \Delta_{i,j}(\Gamma)
\Delta_{i,j}(\Gamma_0)
\label{eqn:ham}
\end{equation}

\noindent where the sum is taken over all pairs of amino acids,
$\Gamma_0$ is the target structure, $\Delta_{i,j}(\Gamma)$ is the
contact map of structure $\Gamma$:
\begin{equation}
\Delta_{ij}(\Gamma) = \{
\begin{array}{l l}
1 &\ \ \ { R_{ij}<r_0 \mbox{ and } |i-j| >2;}\\
0 & \ \ \ \mbox{ otherwise, }
\end{array}
\end{equation}

\noindent where $R_{ij}$ is the distance of amino acids $i$ and $j$.

 The polypeptide chain is modelled as a chain of beads subject to
steric constraints \cite{CJ,19}. We adopted a discrete representation
similar to the one of Covell and Jernigan \cite{CJ,PL95}, in which
each bead occupies a site of an FCC lattice with lattice spacing equal
to 3.8 \AA. Such a representation is able to describe the backbone of
natural proteins to better that 1 \AA \ rmsd per residue (equal to the
best experimental resolution) and preserves typical torsional
angles. All discretized structures were subject to a suitable
constraint: any two non-consecutive residues cannot be closer than
$4.65$ \AA\ due to excluded volume effects and the distance between
consecutive residues can fluctuate between 2.6 \AA $< d <$ 4.7
\AA. Such constraints were determined by an analysis of the
coarse-grainings of several proteins of intermediate length ($\approx
100$ residues). In order to enforce a realistic global compactness for
a backbone of length $L$, the number of contacts in all the target
structures considered was chosen \cite{cont} to be around $N=1.9L$
while, locally, no residue was allowed to make contact with four or
more consecutive residues.

In order to assess the validity of the variational principle, it is
necessary to evaluate the typical time, $t(\Gamma_0)$, taken to fold
into a given target structure, $\Gamma_0$, followed by a selection of
the structures $\Gamma_0$, that have the smallest folding times. To
do this, an initial set of ten conformations was generated by
collapsing a loose chain starting from random initial conditions. In
each case, we modified the random initial conformation by using Monte
Carlo dynamics: we move up to 3 consecutive beads to unoccupied
discrete positions that do not violate any of the physical constraints
and accept the moves according to the standard Metropolis rule. The
energy is given by eq. (\ref{eqn:ham}), while the temperature for the
MC dynamics was set to 0.35. This value was chosen in preliminary runs
so that it was higher than the temperature \cite{Woly} below which the
sequence is trapped in metastable states but comparable to the folding
transition temperature so that conformations with significant overlap
with the native state are sampled in thermal equilibrium.

For each structure, as a measure of the folding time we took the
median over various attempts (typically 41) of the total number of
Monte Carlo moves necessary to form a pre-assigned fraction of native
contacts, typically 66\%, starting from a random conformation. Our
results were unaltered on increasing this fraction to 75\%; indeed,
this fraction could be progressively increased towards 100\% with
successive generations without increase in the computational cost
since better and better folders are obtained.

A new generation of ten structures is created by ``hybridizing'' pairs
of structures of the previous generation ensuring that structures with
small folding times are hybridized more and more frequently as the
number of generations, $g$, increases \cite{14}. To do this, each of
the two distinct parent structures to be paired, $\Gamma_1$ and
$\Gamma_2$ are chosen with probability proportional to
$\exp[-(g-1)*f_t)/1000]$, where $g$ is the index of the current
generation (initially equal to 1), $f_t$ is the median folding
time. Then, a hybrid map is created by taking the union of the two
parent maps:
\begin{equation}
\Delta_{ij}^{Union} = \max(\Delta_{ij}(\Gamma_1),\Delta_{ij}(\Gamma_2))\ .
\end{equation}

\noindent Because it is not guaranteed that $\Delta^{Union}$
corresponds to a three-dimensional structure obeying the same physical
constraints as $\Gamma_1$ and $\Gamma_2$, the corresponding hybrid
$\Gamma$ is constructed by taking one of the two parent structures (or
alternatively a random one) as the starting conformation and carrying
out MC dynamics favouring the formation of each of the contacts in the
union map (i.e. using eq. (\ref{eqn:ham}) with $\Delta_{ij}(\Gamma_0)$
substituted by $\Delta_{ij}^{Union}$). The dynamics is carried out
starting from a temperature of $0.7$ and then decreasing it gradually
over a sufficiently long time (typically thousands of MC steps) to
achieve the maximum possible overlap with the union map, while
simultaneously maintaining the realistic compactness. The resulting
structure is typically midway between the two parent structures, in
that it inherits native contacts from both of them. We adopted the
following definition in order to obtain an objective and unbiased way
to quantitatively estimate the presence of secondary content: a given
residue, $i$ was defined to belong to a secondary motif if, for some
$j$, one of these conditions held:

\begin{eqnarray}
a)\ \Delta_{i-1,j-1}&=&\Delta_{i,j}=\Delta_{i+1,j+1}=\Delta_{i,j+1}\nonumber \\
    &=& \Delta_{i+1,j+2}=\Delta_{i-1,j}=1;\nonumber \\
b)\ \Delta_{i+1,j-1}&=&\Delta_{i,j}=\Delta_{i-1,j+1}=\Delta_{i,j+1}\nonumber \\
    &=& \Delta_{i+1,j}=\Delta_{i-1,j+2}=1.\nonumber
\end{eqnarray}

\noindent The former [latter] identifies the presence of helices and
parallel [anti-parallel] $\beta$ sheets in natural proteins, which can
be identified by the visual inspection of contact matrices and appears
as thick bands parallel or orthogonal to the diagonal.

\begin{figure}
\centerline{\psfig{figure=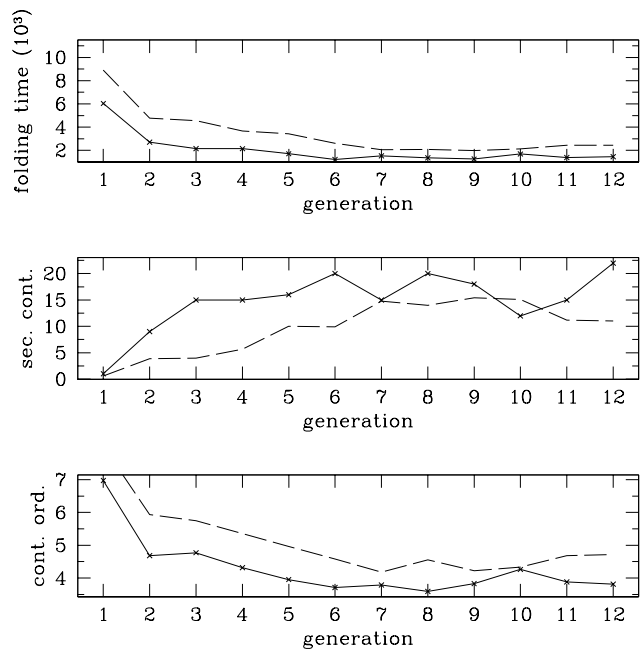,width=3.5in}}
\caption{Evolution of the median folding time (measured in Monte Carlo
steps), secondary structure content and contact order as a function of
the number of generations in the optimization   algorithm for compact
structures of length $L=25$. The dashed curve denotes an average over
all ten structures in a given generation, whereas the solid curve
shows the behaviour of the structure at each generation with the
fastest median folding time. Analogous results are obtained for other
runs and for other values of $L$. The dramatic decrease of folding
time is accompanied by an equally significant increase in the
secondary content.}
\label{fig:fastfig1}
\end{figure}

\noindent The upper plot of Fig. \ref{fig:fastfig1} shows the decrease of
the typical folding time over the generations for chains of length 25,
while the middle panel shows the accompanying increase in the number
of residues in secondary motifs (secondary content). The bottom panel
shows a milder decrease of the contact order (i.e. a larger number of
short-range contacts) as the generations evolved, in agreement with
the experimental findings of Plaxco {\em et al.}\cite{Pl}

\begin{figure}
\centerline{\psfig{figure=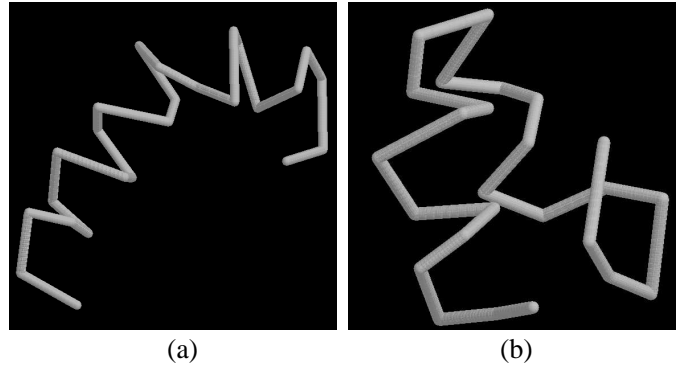,width=3.5in}}
\caption{ a) RASMOL plot of a structure with very low median folding
time and $L=25$. b) Structure with very low median folding time,
$L=25$ and higher compactness (all target conformations were
constrained to have a radius of gyration smaller than 6.5
\AA). Optimal compact structures correspond to helices packed
together, as observed in naturally occurring proteins.}
\label{fig:fastfig2}
\end{figure}

\begin{figure}
\centerline{\psfig{figure=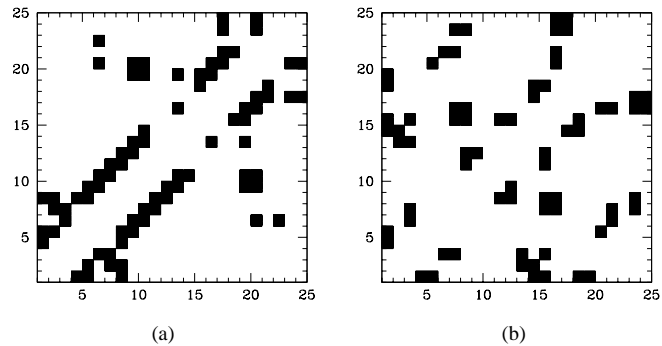,width=3.5in}}
\caption{The panel on the left [right] shows the contact map of a
structure with a very low [average] median folding time. The signature
of helices in map (a) is shown by the thick bands parallel to the
diagonal, while no such patterns are observed in the matrix (b).}
\label{fig:fastfig3}
\end{figure}

One of the optimal structures of length 25 is shown in Figure
\ref{fig:fastfig2}a. Due to the absence of any chirality bias in our
structure space exploration, the helix does not have a constant
handedness. The signature of the secondary motifs in the optimal
structures is clearly visible in the contact maps of Figure
\ref{fig:fastfig3}, which are not sensitive to structure chirality.
Strikingly, the variational principle selects conformations with
significant secondary content as those facilitating the fastest
folding. It is also noteworthy that the average value of $f$ defined
in Sec. \ref{sec:tube} is 0.9, which is very close to the ideal value,
$f=1$, despite the fact that the underlying FCC lattice prevents the
structures from attaining a regular helical shape.

\begin{figure}
\centerline{\psfig{figure=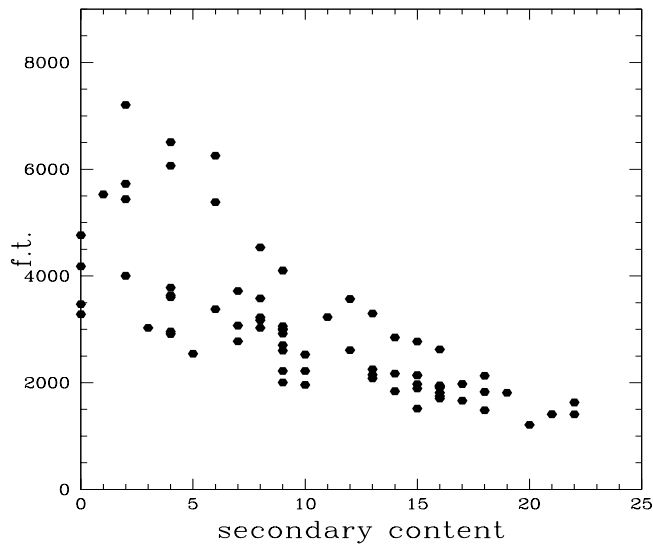,width=3.5in,height=3.0in}}
\caption{Scatter plot of folding time versus secondary content for
structures of length 25 collected over several generation of the
optimization algorithm.}
\label{fig:fasttrend}
\end{figure}

The correlation of the emergence of secondary structures with decrease
of folding times is shown in the plot of Fig. \ref{fig:fasttrend}. We
verified that the hybridization procedure is not biased towards low
contact order by iterating it for various generations and pairing the
structures at random. Even after dozens of generations, the generated
structures had secondary contents of about 1/3-1/4 of the true
extremal structures.

The very high secondary content in optimal conformations was found to
be robust against changes in chain length or compactness of the target
structure. On requiring that the structure be more compact, bundles of
helices emerge [see Fig. \ref{fig:fastfig2}b] along with an increase
in contact order, signalling the presence of some longer range
contacts, which are necessitated in order to accomodate the shorter
radius of gyration. It is noteworthy that our calculations lead
predominantly to $\alpha$-helices and not $\beta$ sheets, a fact
accounted for by the demonstration that steric overlaps and the
associated loss of entropy lead to the destabilization of helices in
favor of sheets \cite{Aurora}, the appearance of such sheets only in
sufficiently long proteins\cite{22} and the much slower folding rate
of $\beta$-sheets compared to $\alpha$-helices \cite{slow}. It is
remarkable that the same requirement of rapid folding is sufficient to
lead to a selection in both sequence and structure space underscoring
the harmony in the evolutionary design of proteins. The results and
strategies presented here ought to be applicable in
protein-engineering contexts, for example by ensuring optimal
dynamical accessibility of the backbone of proteins. A systematic
collection of the rapidly-accessible structures of various length
should also lead to the creation of unbiased libraries of protein
folds.

\section{Density of overlapping conformations for protein
structures and role of native state topology in the folding process}
\label{sec:dos}

The rapid and reversible folding of proteins-like heteropolymers into
their thermodynamically stable native state \cite{Anf} is accompanied
by a huge reduction in conformational entropy \cite{KarplusW,Pt92}.
Evidence has been accumulating for an achievement of the entropy
reduction through a folding funnel favoring the kinetic accessibility
of the native state \cite{due,Woly,LMO,tre,trebis}. In
Sec. \ref{sec:tube} and \ref{sec:fast}, we have seen that secondary
motifs of proteins may arise from the requirements of both being
compact and easily kinetically accessible. Here we focus on a further
characterization of the special role played by the native structure of
proteins; again we will make no use of detailed information regarding
amino acid sequences \cite{19}. The study is carried out through a
theoretical probe of the conformation space of proteins: a measure of
the density of overlapping conformations (DOC) having a given overlap
or percentage of contacts in common with a fixed native structure. We
show with studies on chymotrypsin inhibitor (reference 2ci2 of the
Protein Data Bank) and barnase (1a2p) that the DOC provides key
information on the folding pathway. An analysis of the DOC for real
protein structures and for artificially generated decoy ones suggests
that an extremal principle is operational in nature, which maximizes
the DOC at intermediate overlap, providing a large basin of attraction
\cite{DC,BOSW95,Woly,LMO,tre} for the native state and promoting the
emergence of secondary structures.

 Our study consists of determining the number of structures with a
given structural similarity to a putative native state. The
structural similarity between the native structure and another one is
defined as the percentage of native contacts in the alternative
conformation. It is well known that such a measure is a good
coordinate characterizing the folding process \cite{Go,CT,SSK}. To
this purpose we adopt the Go scheme introduced in the previous section
(see eqn. \ref{eqn:ham}). We also make again use of the FCC
coarse-graining to avoid considering as distinct conformations that
differ slightly. 

The generation of conformations was carried out using a standard Monte
Carlo procedure (see e.g. Refs. \cite{SSK,KS}) which allows one
to move simultaneously up to 7 randomly chosen $C_\alpha$'s to
unoccupied FCC sites.

In order to minimize the effects of correlation between successively
generated structures, we typically discarded 50 elementary moves
before accepting each new conformation. A newly generated
conformation was accepted with the usual Metropolis rule according to
the change in the Boltzmann weight: $e^{\Delta / K_B T}$, where
$\Delta$ is the change in contact overlap and $T$ is a fictitious
temperature. By choosing $T$ appropriately, one can readily generate
conformations with a desired average contact overlap, $\bar{q}$. At a
given temperature, the true number of structures with overlap $q$ is
proportional to the number of conformations with overlap $q$ obtained
in the simulation multiplied by the Boltzmann weight. On undoing the
Boltzmann bias, it is possible to recover the true density of
conformations in a region around $\bar{q}$. In order to obtain the
density of conformations for all values of overlap, we performed {\em
2500} Monte Carlo samplings at different decreasing temperatures and
then used standard deconvolution procedures \cite{FS89}. Overall,
for each distinct value of the overlap, more than 1000 structures were
sampled. We have confirmed that the DOC curves are independent of the
starting conformation and that the ``folding'' DOC obtained starting
from a random conformation and cooling agrees to better than 3 \% with
the ``unfolding'' DOC obtained starting from the target structure and
increasing the temperature.

\begin{figure} 
\centerline{\psfig{figure=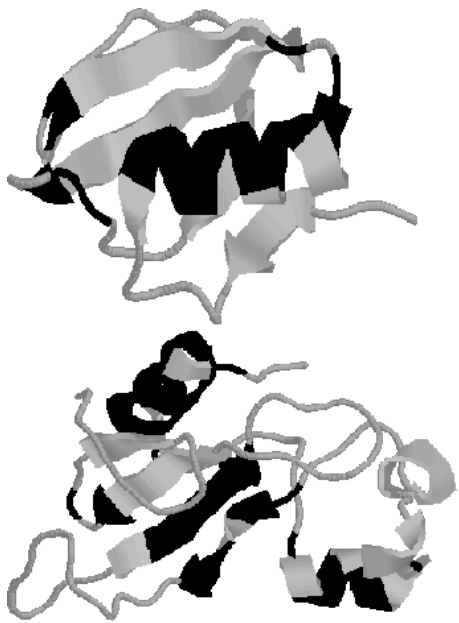,height=3.2in}} \caption{Ribbon plot
(obtained with RASMOL) of 2ci2 (top) and barnase (bottom). The residues
involved in the 12 [16] most frequent contacts of alternative structures with
overlap $\approx 40 \%$ with the native conformations are highlighted in
black. The majority of these coincide with contacts that are formed at the
early stages of folding.} \label{fig:dos2ci2} \end{figure}

\begin{figure}
\centerline{\psfig{figure=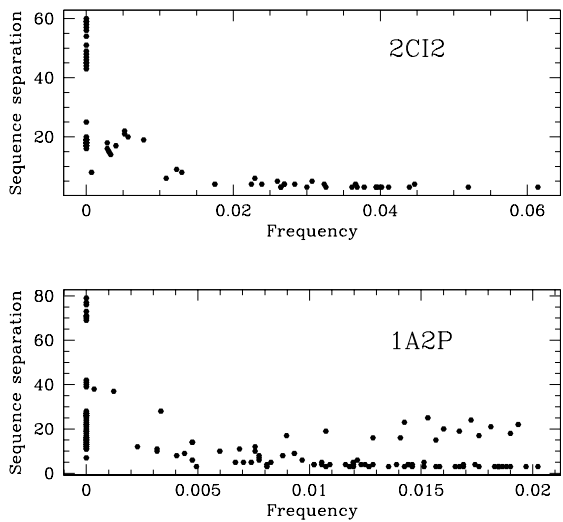,width=3.1in}}
\caption{Distribution of sequence separation of contacts commonly found in
the conformations that overlap with the native state structures of
2ci2 and 1a2p. The most frequent contacts for 2ci2 have
a small sequence separation (3-4) and pertain to helix formation.
The 1a2p case shows a very different
behaviour with several contacts with very large sequence separation. }
\label{fig:dosfsep}
\end{figure}

We begin with the backbones of the chymotrypsin inhibitor and barnase.
We generated 2500 structures with a not too large overlap
\cite{footnote1} ($\approx 40 \%$) for each of them. It turned out
that the most frequent contacts shared by the native conformation of
2ci2 with the others involved the helical-residues 30-42 (see top
Fig. \ref{fig:dos2ci2}). Contacts involving such residues were shared by
$56 \%$ of the sampled structures. On the other hand, the rarest
contacts pertained to interaction between the helix and
$\beta$-strands and between the $\beta$-strands themselves. A
different behaviour (see bottom Fig. \ref{fig:dosfsep}) was found for
barnase, where, again, for overlap of $\approx 40 \%$, we find many
contacts pertaining to the nearly complete formation of helix 1
(residues 8-18), a partial formation of helix 2, and bonds between
residues 26-29 and 29-32 as well as several non-local contacts
bridging the $\beta$-strands, especially residues 51-55 and
72-75.

Both this picture, and the one described for CI2 are fully consistent
with the experimental results obtained by Fersht and co-workers in
mutagenesis experiments \cite{F95,ci2b}. In such experiments, the key
role of an amino acid at a given site is probed by mutating it and
measuring the changes in the folding and equilibrium characteristics.
By measuring the change of the folding/unfolding equilibrium constant
one can introduce a parameter, termed $\phi$-value, which is zero if
the mutation is irrelevant to the folding kinetics and 1, if the
change in folding propensity mirrors the change in the relative
stability of the folded and unfolded states (intermediate values are,
of course, possible). Ideally, the measure of the sensitivity to a
given site should be measured as a suitable susceptibility to a small
perturbation of the same site (or its environment). Unfortunately,
this is not easily accomplished experimentally, since substitution by
mutation can be rarely regarded as a perturbation. Notwithstanding
this difficulty, from the analysis of the $\phi$-values obtained by
Fersht, a clear picture for the folding stages of CI2 and barnase
emerges. In both cases, the crucial regions for both proteins are the
same as those identified through the analysis of the DOC reported
above. This provides a sound {\em a posteriori}\/ justification that
the main features of the folding of a protein can be followed from a
study of the DOC. Remarkably, the method discussed above relies
entirely on structure-related properties and suggests that the main
features of the folding funnel are determined by the geometry of the
``bare'' backbone, while the finer details, of course, depend on the
specific well-designed sequence. Since our own work \cite{19}, other
groups have used similar or alternative techniques to elucidate the
role of the native state topology in the folding process
\cite{Finkel,slow,baker,Clem}, confirming the picture outlined here.

Let us consider one way in which 
proteins, in general, are special and different from arbitrary compact
polymers. To do so, we turn to an analysis of three proteins of length
51 (1hcg, 1hja and 1sgp) which have nearly the same number of native
contacts ($\approx 83$). For each structure, we calculated the DOC
with the constraint that the total number of contacts in the
alternative structures do not exceed the number of contacts in the
native state by more than 10 \% to avoid excessive compactness.
To assess whether the DOC associated with naturally occurring proteins
have special features, we generated three decoy conformations of the
same length and number of contacts, but with different degrees of
short and long range contacts (in sequence separation). These decoys
(subject to the aforementioned ``physical constraints'') were
generated with a simulated annealing procedure to find the structure
with the highest overlap with a target contact matrix. By tuning the
number of short-range versus long-range entries in the target random
contact matrix, we generated three structures with different degrees of
compactness and local geometrical regularity.

\begin{figure}
\centerline{\psfig{figure=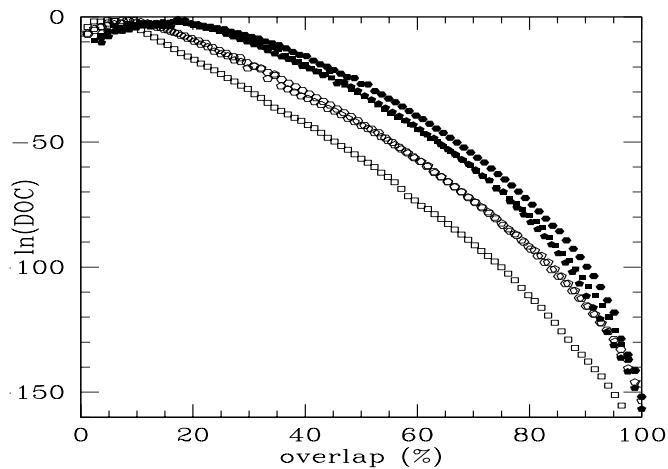,height=2.5in,width=3.5in}}
\caption{Density of overlapping conformations
for proteins for 1sgp (filled squares),
1hja (filled pentagons) and 1hcg (filled hexagons). Curves for
artificial decoy structures are denoted by the open symbols.}
\label{fig:dosdos1}
\end{figure}

The logarithmic plots of the DOC are shown in Fig. \ref{fig:dosdos1}. A
striking feature of the curves is that, for intermediate overlap, the
DOC of the real proteins is enormously larger than that of the decoys
and suggests that naturally occuring conformations have a much larger
number of entryway structures than random compact
conformations. Furthermore, for very high values of the overlap, the
steepness of the protein curves is much larger than those of the
decoys, showing that the reduction in the conformational entropy is
also higher. This implies the existence of a funnel with a very large
basin and steep walls. Another significant feature is the good
collapse of the protein curves. We verified that this feature also
obtains for 1bd0 and 2pk4 which each have 80 residues and 140 and 146
contacts respectively. A simple explanation for the curve collapse,
could be that the DOC of real proteins is ``extremal'', in that it is
close to the maximum possible value for intermediate values of the
overlap.

The importance of the locality of contacts for folding kinetics was
highlighted recently by Plaxco {\em et al.} \cite{Pl} who found a
correlation between folding rate and contact order, defined as the
average sequence separation of contacts normalized to the total number
of contacts and sequence length. With reference to
Fig. \ref{fig:dosdos1}, the contact order values for proteins 1hcg, 1hja
and 1sgp are 0.139, 0.214 and 0.204 respectively. For the decoy
structures, they are 0.424, 0.222 and 0.179 for the curves denoted by
open squares, pentagons and hexagons, respectively. The structure
with an unusually high contact order has the lowest DOC curve and
optimal sequences designed on it (or equivalently a Go-like model)
would be expected to exhibit slow folding dynamics \cite{PGW} in
accord with the findings of Plaxco {\em et al.} \cite{Pl}.

Secondary structure motifs \cite{Creighton,Li}  have characteristic
signatures in the contact maps, such as bands parallel to the diagonal
($\alpha$-helices and parallel $\beta$-sheets) or orthogonal to it
(antiparallel $\beta$-sheets), as it has been shown in the previous
section.
 We have carried out some simple
investigations to assess whether a correlation exists between the
extremality of the DOC curve and the emergence of
secondary-structure-like motifs. We considered a space of contact maps
\cite{Levit}, within which each of the residues interacted with the
same number of other residues, $n_c$ (typically $n_c=5$, as in the
average case of a protein with about 100 residues and a cutoff
distance of 6.5 \AA). This space contains maps corresponding to
both real structures and unphysical ones. Furthermore, to mimic the
effects of the rigidity and geometry of the peptide bond, we
disallowed contacts between residue $i$ and the four neighboring
residues along the sequence $i\pm 2$, $i\pm 1$.

 In this context, the maximization of the density of states
corresponds to finding the target matrix with the highest number of
matrices sharing a given fraction of its contacts. Although it is
difficult to solve this problem, for arbitrary values of the overlap,
it is relatively easy to generate matrices with an overlap close to
the maximum value, $\bar{q}_{max}$ (for a $L$x$L$ matrix,
$\bar{q}_{{max}}=L\cdot n_c$). To enumerate all matrices with overlap
$\bar{q}_{{max}}-2$, one first identifies a pair of non-zero entries
in the target matrix ${\bar m}$: $\bar{m}_{ij}=\bar{m}_{kl}=1$. Then
it is necessary to check whether entries $\bar{m}_{il},\bar{m}_{kj}$
are both ``free'' (i.e. equal to zero) and do not correspond to
forbidden contacts (e.g. between $i$ and $i+1$). If this is so, the
old pair of entries (and their symmetric counterpart) are set to zero,
and the new ones to 1. By considering, in turn, all possible pairs of
non-zero entries one can generate all matrices of overlap
$\bar{q}_{{max}}-2$ . Then, by performing a simulated annealing in
contact-map space one can isolate the map having the highest number of
matrices with overlap $\bar{q}_{max}-2$.

\begin{figure}
\centerline{\psfig{figure=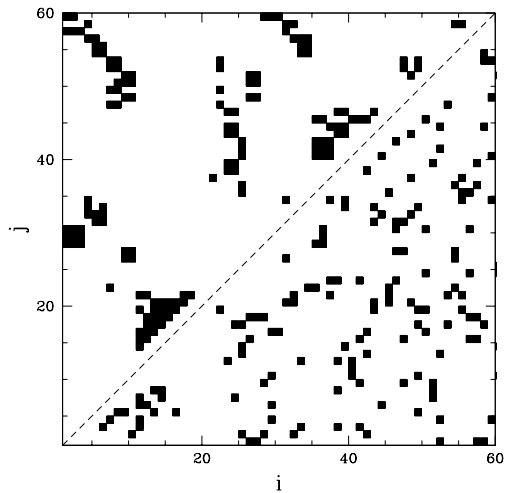,width=2.7in}}
\caption{The upper [lower] triangle shows a target contact matrix with
$L=60$ that has a large [intermediate] number of contact maps with an
overlap of $\bar{q}_{{max}}-2$ contacts.}
\label{fig:dosmap}
\end{figure}

\begin{figure}
\centerline{\psfig{figure=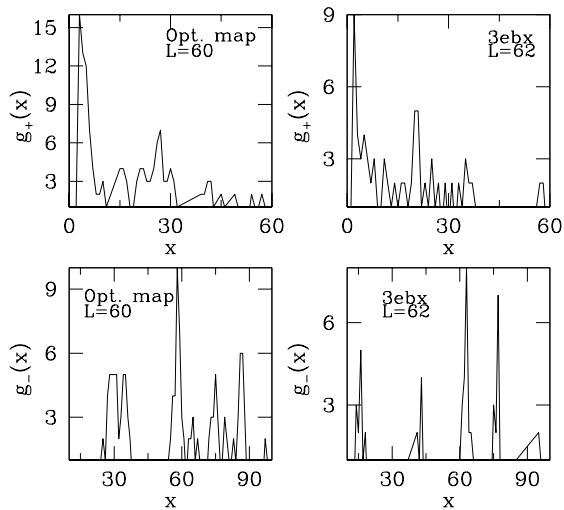,width=3.0in}}
\caption{Correlation functions (see text) for an
optimal target matrix of length 60 and for a protein of length 62
taken from the protein data bank.}
\label{fig:doscorr}
\end{figure}

We carried out our calculations for values of $L$ around 60. The
optimal matrices exhibit clustering reminiscent of $\alpha$-helices
and $\beta$-sheets, as shown in the upper triangle of
Fig. \ref{fig:dosmap}. A more quantitative measurement of the
secondary-structure content of the optimal matrices can be obtained by
considering the correlation functions, $g_{\pm} (x) = \sum_{i} m_{i,i
\pm x}$, which show peaks in correspondence with the sequence
separation of residues involved in $\alpha$-helices and parallel
$\beta$-sheets ($g_{+}$) or antiparallel $\beta$-sheets ($g_{-}$). A
typical plot of the correlation functions for an optimal map of length
60 and for the protein 3ebx (length 62) are shown in
Fig. \ref{fig:doscorr}. The similarity of the plots is striking,
particularly because, in both cases, the height of the peaks in
$g_{+}$ decreases with sequence separation, unlike the situation with
$g_{-}$.

In summary, the geometry of protein backbones seems to have been
optimized to provide a large basin of attraction to the native
state. The results presented here are suggestive of an extremality
principle underlying the selection of naturally occurring folds of
proteins which, in turn, is shown to be possibly associated with the
emergence of secondary structures. This observation complements the
one made in the previous section, which highlighted how the presence
of secondary motifs boosts the folding kinetics. Strikingly, by
examining the DOC associated with a given native structure, it is
possible to extract a wealth of information about the sites involved
in crucial stages of the folding process. Despite the fact, that such
sites are determined from the analysis of their crucial topological
role with respect to the native state with no input of the actual
protein composition, they correlate very well with the key sites
determined experimentally. A striking example is provided in the
following subsection, which focuses on an enzyme encoded by the HIV
virus. It will be shown that from the mere knowledge of the
contact map of the enzyme, one can isolate a handful of hot sites
which correlate extremely well the key mutating sites determined in
clinical trials of anti-AIDS drugs.

\subsection{Application to HIV-1 protease: drug resistance and folding
Pathways}
\label{sec:hiv}

\begin{figure}
\centerline{\psfig{figure=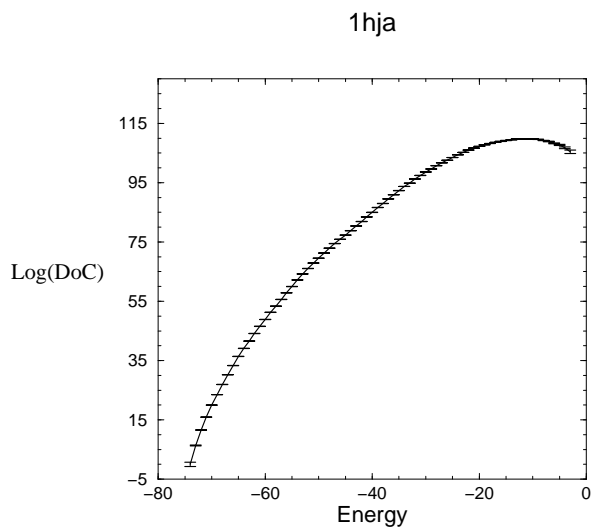,width=3.0in}}
\caption{Plot of the logarithm of the density of states (density of
alternative conformations) versus energy for protein 1hja. The latter
is defined as the negative of the number of native contacts present in
a given conformation.}
\label{fig:hivfig1}
\end{figure}

\begin{figure}
\centerline{\psfig{figure=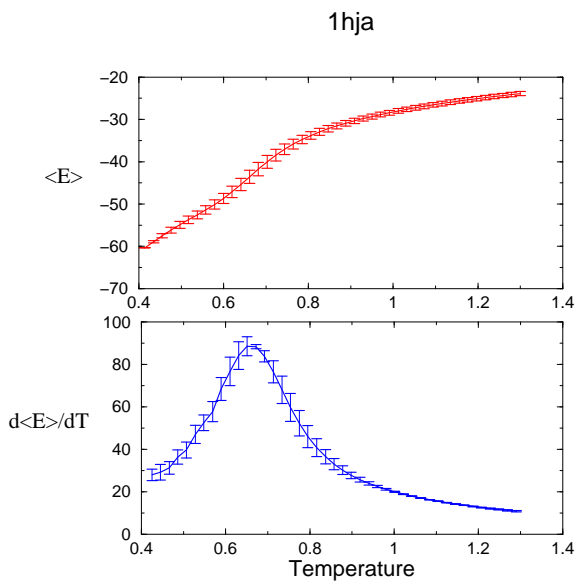,width=3.0in}}
\caption{Plots of the energy (top) and specific heat (bottom) as a
function of temperature for protein 1hja. The curves were obtained
through histogram reweighting techniques.}
\label{fig:hivfig2}
\end{figure}

A close examination of the curves for the density of states (DOC) can
reveal the presence of phase transitions in the system (or more
accurately, their analog for finite-size systems). In particular, a
change of the concavity/convexity of the DOC curves signals a
first-order like transition. This is illustrated in the examples of
Figs. \ref{fig:hivfig1} and \ref{fig:hivfig2} where the curve
for the DOC of protein 1hja is shown and accompanied with a graph of
the energy versus temperature and the specific heat (obtained through
histogram reweighting techniques). Indeed, the peak in the specific
heat identifies the folding transition temperature. Naturally, one
would like to identify the pairs of sites that contribute most to the
$C_v$ peak. For a given pair, $(i,j)$, this amounts to measuring, as
a function of the temperature, $T$, the quantity

\begin{equation}
C_v(i,j) = {1 \over T^2} \langle E(i,j) E_{Tot} \rangle - \langle E(i,j)
\rangle \langle E_{Tot} \rangle
\end{equation}

\noindent where $E(i,j)$ is the contribution of the pair to the total
interaction energy, $E_{Tot}$. The most crucial residues will then be
identified with those with the highest values of $C_v(i,j)$, which are
expectedly observed near the folding transition temperature. We have used this
strategy to determine succesfully the key sites of HIV-1 \cite{aids}.
Our simulations, at variance with that described earlier, are now carried out
in the continuum. As usual, amino acids are represented as effective
centroids placed on $C_\alpha$ atoms, while the peptide bond between
two consecutive amino acids, $i$, $i+1$ at distance $r_{i,i+1}$ is
described by the anharmonic potential adopted by Clementi et al. \cite{CCM}, with
parameters $a = 20$, $b = 2000$. The interaction among
non-consecutive residues is treated again in Go-like schemes \cite{Go}
which reward the formation of native contacts with a decrease of the
energy scoring function. Each pair of non-consecutive amino acids,
$i$ and $j$, contributes to the energy scoring function by an amount:

\begin{equation}
V_0 \varepsilon_{ij}^N \left[ 
\left( 5 \frac{r^N_{ij}}{r_{ij}} \right)^{12} - 6
\left( \frac{r^N_{ij}}{r_{ij}} \right)^{10} \right] + 
5 V_1 (1-\varepsilon^N_{ij}) \left(\frac{r_0}{r_{ij}}\right)^{12},
\end{equation} 

where $r_0 = 6.8$\AA, $r^N_{i,j}$ denotes the distance of amino acids
$i$ and $j$ in the native structure and $\varepsilon^N_{ij}$ is the
native contact matrix whose entries are 1 (0) if $i$ and $j$ are (not)
in contact in the native conformation. The contact is defined as in
the previous section, i.e. two aminoacids are in contact if their
distance is less than $6.5$ {\AA}. $V_0$ and $V_1$ are constants
controlling the strength of interactions ($V_0 = 20$, $V_1=0.05$ in
our simulations)

Constant temperature molecular dynamics simulations were carried out
where the equations of motion are integrated by a velocity-Verlet
algorithm combined with the standard Gaussian isokinetic scheme
\cite{Therm,aids}. Unfolding processes can be studied within the same
framework by warming up starting from the native conformation (heat
denaturation).

The free-energy, the total specific-heat, $C_v$, and contributions of
the individual contacts to $C_v$ were obtained combining data sampled
at different equilibrium temperatures with multiple histogram
techniques \cite{FS89}. The thermodynamics quantities obtained
through such deconvolution procedures did not depend, within the
numerical accuracy, on whether unfolding or refolding paths were
followed.

The contacts that contribute more to the specific heat peak are
identified as the key ones belonging to the folding bottleneck and
sites sharing them as the most likely to be sensitive to mutations.
Furthermore, by following several individual folding trajectories (by
suddenly quenching unfolded conformations below the folding transition
temperature, $T_{fold}$) we ascertained that all such dynamical
pathways encountered the same kinetic bottlenecks determined as above.

For the $\beta$ sheets, the bottlenecks involve amino acids that are
typically 3-4 residues away from the turns -- specifically, residues
61, 62, 72, 74 for $\beta_3$, 10, 11, 12, 21, 22, 23 for $\beta_1$ and
44, 45, 46, 55, 56, 57 for $\beta_2$. At the folding transition temperature,
$T_{fold}$, the formation of contacts around residues 30 and 86 is
observed. The largest contribution to the specific heat peak is
observed from contacts 29-86 and 32-76 which are,
consequently, identified as the most crucial for the folding/unfolding
process.

Such sites are physically located at the active site of HIV-1 PR,
which is targeted by anti AIDS drugs\cite{condra1}. Hence, within
the limitations of our simplified approach, we predict that changes in
the detailed chemistry at the active site also ruin a key step of the
folding process. To counteract the drug action, the virus has to
perform some very delicate mutations at the the key
sites; within a random mutation scheme this requires many trials
(occurring over several months). The time required to synthesize a
mutated protein with native-like activity is even longer if the drug
attack correlates with several bottlenecks simultaneously.

This is certainly the case for several anti-AIDS drugs. Indeed Table
\ref{tab:tab} summarizes the mutations for the FDA approved
drugs \cite{BIOCH88}. In Table \ref{tab:tab2}, we list the sites
sharing the most important contacts involved in the four bottlenecks
TB, $\beta_1$, $\beta_2$ and $\beta_3$. Remarkably, among the first 23
most crucial sites predicted by our method, there are 7 sites in common
with the 16 distinct mutating sites of Table \ref{tab:tab}. The
probability that two sets of 16 and 23 sites randomly taken from a
total population of 99 (the length of the HIV-1 PR monomer) share at
least 7 sites is only 3 \%. Also note that, all the mutation sites of
Table \ref{tab:tab} except 82, 35, 36 and 90 fall within a mismatch of
at most one position from the sites of Table \ref{tab:tab2}. These
results highlight the high statistical correlation between our
prediction and the evidence accumulated from clinical trials.

In conclusion, the strategy presented here, which is entirely 
based on the knowledge of the native structure of HIV-1 protease,
allows one both to identify the bottlenecks of the folding process and to
explain their highly significant match with known mutating residues
\cite{aids}. This and similar approaches should be readily applicable
to identify the kinetic bottlenecks of other viral enzymes of
pharmaceutical interest. This could allow a fast development of novel
inhibitors targetting the kinetic bottlenecks. This is expected to
dramatically enhance the difficulty for the virus to express mutated
proteins which still fold efficiently into the same native state with
unaltered functionality.

\begin{table}
\begin{center}
\begin{minipage}{12.0cm}
\begin{tabular}{|l|l|l|} 
\hline
Name & Point Mutations & Bottlenecks \\ 
\hline \\
RTN \protect\cite{Molla,Marko} & {\bf 20}, { 33}, 35, 36, {\bf
46}, 54, {\bf 63}, 71, 82, {\bf 84}, 90 & TB, $\beta_1$, $\beta_2$, $\beta_3$ \\
NLF \protect\cite{patick} & 30, {\bf 46}, {\bf 63}, 71, { 77}, {\bf 84}, 
& TB, $\beta_2$, $\beta_3$ \\
IND \protect\cite{condra,tisdale} & {\bf 10}, {\bf 32}, {\bf 46}, {\bf 63},71, 82,
{\bf 84} & TB, $\beta_1$, $\beta_2$, $\beta_3$ \\
SQV \protect\cite{condra,tisdale,jacob} & {\bf 10}, {\bf 46}, 48, {\bf 63}, 
71, 82, {\bf 84}, 90 & TB, $\beta_1$, $\beta_2$, $\beta_3$ \\
APR \protect\cite{apr} & {\bf 46}, {\bf 63}, 82, {\bf 84} & TB,
$\beta_2$, $\beta_3$ \\
\hline
\end{tabular}
\end{minipage}
\end{center}
\caption{Mutations in the protease associated with FDA-approved drug
resistance \protect\cite{BIOCH88}. Sites highlighted in boldface are
those involved in the folding bottlenecks as predicted by our
approach. $\beta_i$ refers to the bottleneck associated with the
formation of the $i$-th $\beta$-sheet, whereas TB refers to the
bottleneck occurring at the folding transition temperature $T_{fold}$
(see next Table).}
\label{tab:tab}
\end{table}

\begin{table}
\begin{center}
\begin{minipage}{7.0cm}
\begin{tabular}{|l|l|}
\hline
Bottleneck & Key sites \\ \hline 
TB & 22, 29, 32, 76, 84, 86 \\
$\beta_1$ & 10,11,13,20,21,23 \\
$\beta_2$ & 44,45,46,55,56,57 \\
$\beta_3$ & 61,62,63,72,74 \\
\hline
\end{tabular}
\end{minipage}
\end{center}
\caption{Key sites for the four bottlenecks. For each bottleneck, only the
sites in the top three pairs of contacts have been reported.}
\label{tab:tab2}
\end{table}


\section{Application of geometrical models to investigate the folding
of membrane proteins} 
\label{sec:mp}

While the behavior of small water soluble globular proteins is
reasonably well understood\cite{Fersht3,Karplus}, much less is known
about proteins (membrane proteins:
MP)\cite{white1,OM97,vH96,booth,PE90,Papu,MS1,MS2,JW87,Rose88,preprint,
bonaccio} that cross biological
membranes and that control solute transport, signal transmission and
energy conversion between the two sides of the membrane. This lack of
knowledge is related to the difficulty in experimental
handling. Membranes consist of phospholipid bilayers with a
hydrophobic interior: the surface of MP that interact with such an
apolar environment is also hydrophobic and this property causes MP to
aggregate in aqueous solution, unless detergents are used. This
circumstance makes crystallization of MP's diffiuclt and native
structures have been determined only for an handful of them.

Transmembrane proteins (TMP) are the most important and best studied
class of MP \cite{white1,OM97,Biggin}. They are characterized by the
presence of long segments ($20 - 30$) of amino acids with a high
degree of hydrophobicity. In the native structure, these correspond to
the transmembrane segments which are inserted in the lipidic interior
of the membrane \cite{goto}. These segments are predominantly made up
of $\alpha$-helices and $\beta$-sheets. The stability of
$\alpha$-helices and $\beta$-sheets inside the membrane follow from
the formation of hydrogen bonds between the backbone atoms -- other
possibilities are excluded within the apolar
enviroment\cite{white1,PE90}.

A detailed study of TMP has not yet been possible because little is
known about the amino-acids interactions among themselves, with the
solvent and in particular with the lipidic interior of the membrane.
Here \cite{OSBLM}, we present a simple strategy to decipher the folding
kinetics of transmembrane proteins which is directly inspired by the
geometrical approaches we have previously applied to globular
proteins. This approach by-passes the details of the complex
interactions of the protein in the lipid enviroment by introducing
effective potentials, induced by the presence of the membrane and the
associated interface region, that stabilize the native state
structure.
 
Due to the small number of degrees of freedom involved in our scheme, 
the dynamics of the
system can be simulated for the full folding process. Moreover, the free
energies of the most relevant intermediate states and free energy profiles 
along the reaction paths connecting them can be explicitly calculated by
thermodynamic integration.
Thus the model is able to quantitatively discriminate between the possible
reaction paths envisaged for the insertion process of TMP across the
membrane\cite{white1}.

\begin{figure}[tbp]
\inseps{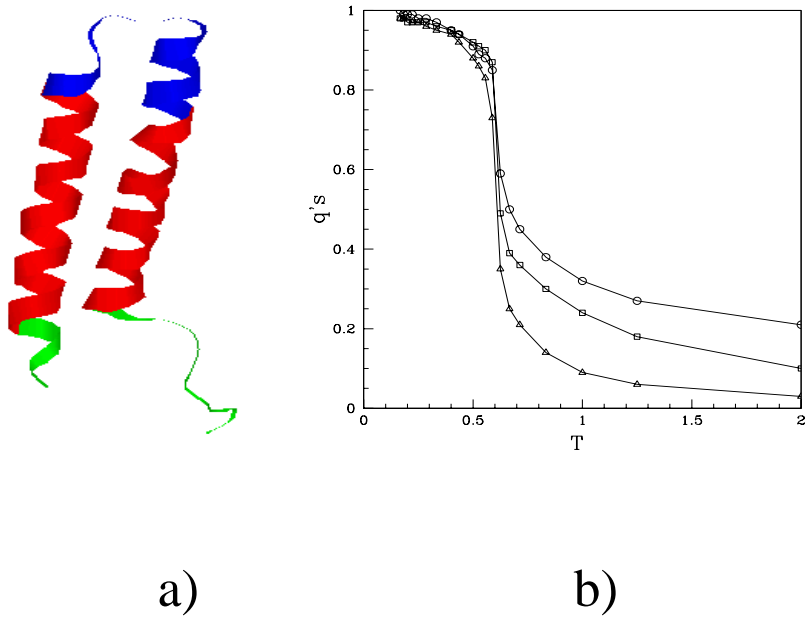}{0.7}
\caption{Structure and thermodynamics of the helical transmembrane
protein.{\bf a)} Ribbon representation of the two-helix fragment of
bacteriorhodopsin formed by the first $66$ amino-acids. The part
inside the membrane is shown in red, the part above (below) the
membrane in blue (green). {\bf b)} Average equilibrium fraction of
native contacts outside, $q_b$ ($\circ$), inside, $q_m$ (square), and
across, $q_s$ ($\triangle$), the membrane as a function of the
temperature $T$. All these quantities are expressed in energy unit of
$\epsilon$.}
\label{fig:mpfig1}
\end{figure}

The TMP we considered is made up with the first 66 amino acids (each
one represented by a fictitious residue located at the position of the
$C_{\alpha}$ atom) of the first two $\alpha $-helices of
bacteriorhodopsin (Fig. \ref{fig:mpfig1}a). It has been shown that
the first two helices of bacteriorhodopsin can be considered as
independent folding domains\cite{KSE92} and that the side-by-side
interactions between transmembrane helices play a key role in the
stabilization of the protein structure\cite{KE92}. The membrane is
described simply by a slab of width $w = z_{{\rm max}}-z_{{\rm min}}=
26$ {\AA}. Two non-bonded residues $(i,j)$ are considered to form a
contact if their distance is less then $ 6.5$ {\AA}. In the study of
globular proteins, the topology of the native state is encoded in the
contact map by considering all the pairs $(i,j)$ of non-consecutive
residues that are in contact. Here, in addition, the locations of such
pairs with respect to the membrane has to be taken into account. The
contacts are divided into three classes:
\begin{itemize}

\item {\sl membrane contacts} where both $i$ and $j$
residues are inside the membrane;

\item {\sl interface contacts} with $i$ and $j$ in
the interface region \cite{white1} outside the
membrane; 

\item {\sl surface contacts} with one residue inside the membrane and
the other outside. 

\end{itemize}

\noindent Thus a given protein conformation can have a native contact
but improperly placed with respect to the membrane ({\sl misplaced
native contact}). The crucial interaction potential between non-bonded
residues $(i,j)$ is taken to be a modified Lennard-Jones 12-10
potential:
\begin{equation}
\Gamma(i,j) \left [ 5\left ( \frac{r^{N}_{ij}}{r_{ij}}\right)^{12}- 
6\left( 
\frac{r^N_{ij}}{r_{ij}}\right)^{10}\right ] +
5\ \Gamma_1(i,j)\left ( \frac{r^{N}_{ij}}{r_{ij}}\right)^{12}.
\end{equation}
\noindent $r_{ij}$ and $r^N_{ij}$ are the distance between the
residues $(i,j)$ and their distance in the native configuration,
respectively. The matrices $\Gamma(i,j)$ and $\Gamma_1(i,j)$ encode
the topology of the TMP in the following way: if $(i,j)$ is not a
contact in the native state $\Gamma(i,j)=0,\Gamma_1(i,j)=1$; if
$(i,j)$ is a contact in the native state but not at the proper
location (i.e. a misplaced contact)
$\Gamma(i,j)=\epsilon_1,\Gamma_1(i,j)=0$; if $(i,j)$ is a native state
contact in the proper region $\Gamma(i,j)=\epsilon,\
\Gamma_1(i,j)=0$. This model is intended to describe the folding
process in the interface and in the membrane region. Our interaction
potential (similar in spirit to the Go model\cite{Go} introduced
before) assigns two values to the energy associated with the formation
of a native contact, $\epsilon$ and $\epsilon_1 $. The model captures
the tendency to form native contacts. In addition, in order to
account for the effective interactions between the membrane and the
protein, the model assigns a lower energy, $-\epsilon$, to the contact
which occurs in the same region as in the native state structure
compared to $-\epsilon_1$ when the contact is formed but in the wrong
enviroment. This mechanism proves to be crucial in driving the
insertion of the protein across the membrane.

When $\epsilon=\epsilon_1$, the protein does not recognize the
presence of the interface-membrane region and the full rotational
symmetry is restored. The difference in the parameters
($\epsilon-\epsilon_1$) determine the amount of tertiary structure
formation outside the membrane. Our results are independent of the
precise values of the energy parameters $\epsilon$ and $ \epsilon_1$
($\epsilon > \epsilon_1$) as long as they are not too close to each
other.

Our simulations have been performed with $\epsilon_1 = 0.1$ and
$\epsilon = 1$. In order to account for the chirality of the TMP, a
potential for the pseudodihedral angle $\alpha_i$ between the
$C_{\alpha}$ atoms in a helix corresponding to four successive
locations is added which biases the helices to be in their native
state structure.

The thermodynamics and the kinetics of the model were studied by a
Monte Carlo method for polymer chains carried out in the continuum.
 The efficiency of the program (usually low for continuum
calculations) has been increased by full use of the link cell
technique \cite{Binder} and by the multiple Markov chain method, a new
sampling scheme, which has been proven to be particulary efficient in
exploring the low temperature phase diagram for
polymers\cite{TJOW}. In our simulation $20$ different temperatures
(measured in dimensionless units) ranging from $T=2$ to $T=0.17$ have
been studied.

The free energy difference ${\cal F}_B-{\cal F}_A$ between two states
A and B has been estimated as the reversible work that has to be done
in order to go from A to B\cite{OSBLM}. The free energy differences
obtained with this method are accurate to within $ \sim $ 0.1$\,T_C$
for the various states whereas the free energy barriers are accurate
within $\sim $ 0.5$\,T_C$ . This error takes into account possible
hysteresis effects due to the finite simulation time.

The structural similarity between the system equilibrated at
temperature $T$ and the native state is shown in Figure
\ref{fig:mpfig1}b in terms of the average fraction of native state
contacts as a function of $T$ and partitioned depending on their
positions with respect to the membrane. The three curves correspond
respectively to the average fraction of native contacts inside ($q_m$)
, outside ($q_b$) and across ($q_s$) the membrane. All these curves,
well separated at high T, collapse for $T$ below the transition
temperature $T_C \sim 0.6$, indicating a cooperative effect in the
folding. On monitoring the free energy as a function of the energy
around $T_C$, one observes additional local minima (besides those
corresponding to the unfolded and folded states) suggesting the
presence of an intermediate.

\begin{figure}[tbp]
\inseps{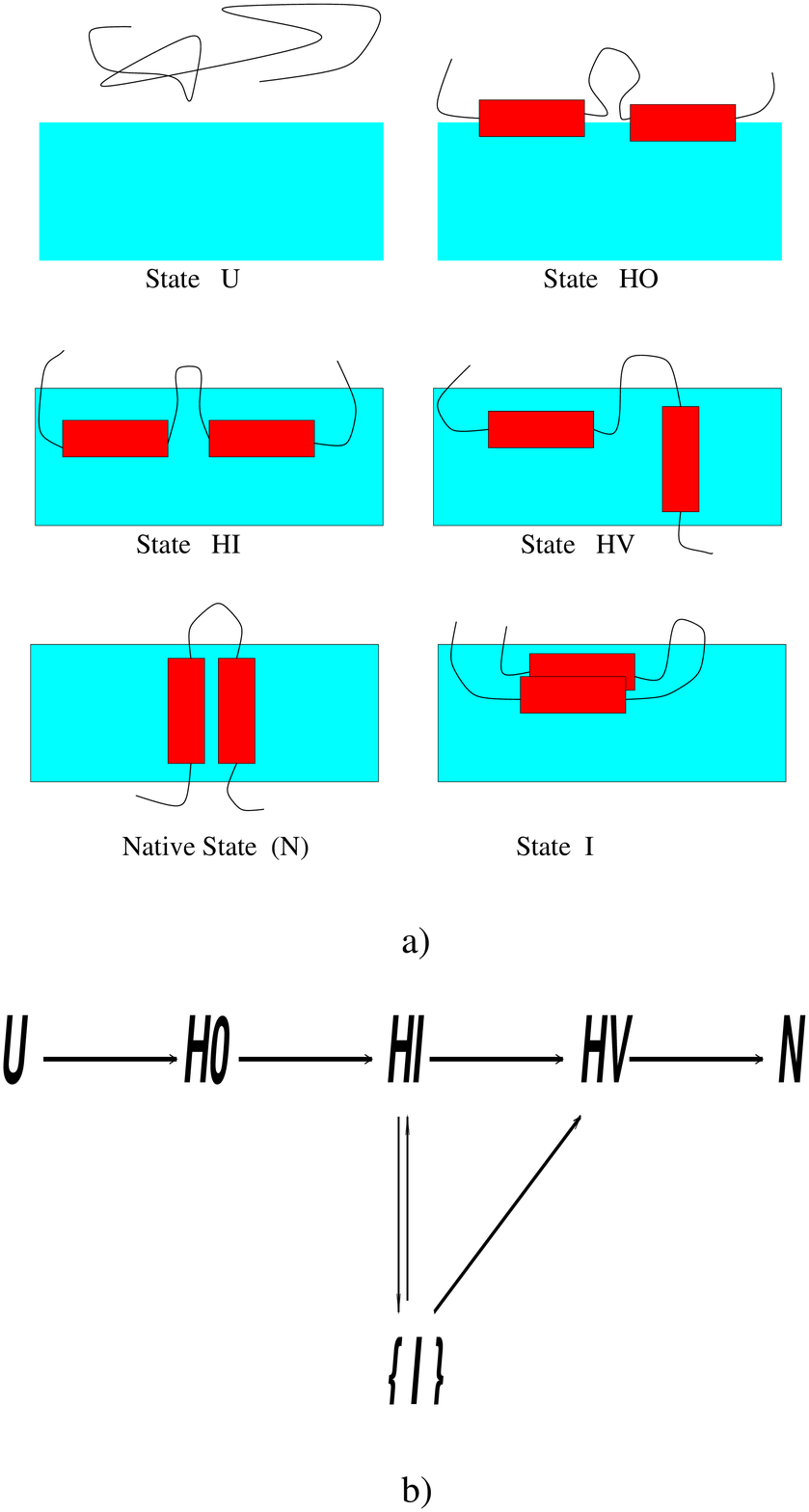}{0.40}
\caption{Schematic representation of states encountered by our model
protein during the folding process. In {\bf a)} the red cylinders
denote $\alpha$-helices that reside within the membrane in the native
state. The region inside the membrane is in turquoise whereas the rest
represents the interface region \cite{white1}. State $U$ denotes the
denatured state of the protein, $HO$ is a state in which the helices
have been formed but are not yet inside the membrane whereas $HI$
corresponds to a similar state but with the helices completely
embedded in the membrane without any inter-helical contacts. $HV$
denotes an obligatory intermediate and $N$ depicts the native state.
The state $\{ I \}$ represents an ensemble of long lived conformations
in which helices are formed inside the membrane with several
inter-helical contacts, but with the two $\alpha$-helices still
incorrectly positioned. In {\bf b)} the schematic pathways to the
native state are shown.}
\label{fig:mpfig2}
\end{figure}

The intermediate is characterized by having the two helices almost
completely formed but not yet correctly inserted across the membrane.
The presence of these extra minima suggests that non-constitutive
membrane proteins would fold with multi-state kinetics corresponding
to on-pathway intermediates. To establish the nature of the dominant
folding pathway, we have performed a detailed analysis of the folding
kinetics. Each independent kinetic folding simulation was started with
the equilibrated denaturated state at $T^* = 2.5$ . The protein is
placed initially outside the membrane in the interface region
\cite{white1}, at a distance comparable to the average size of the
denatured protein and then suddenly quenched to a temperature
($T=0.4$) well below the transition temperature. This case simulates
the folding kinetics of non-constitutive membrane proteins,
i.e. proteins that do not need a translocon providing a 'tunnel'
through which the protein is injected into the lipid bilayer. Folding
to the native state occurs mainly through the states depicted in
Figure \ref{fig:mpfig2}a with the dominant pathways shown in Figure
\ref{fig:mpfig2}b.

\begin{figure}[tbp]
\inseps{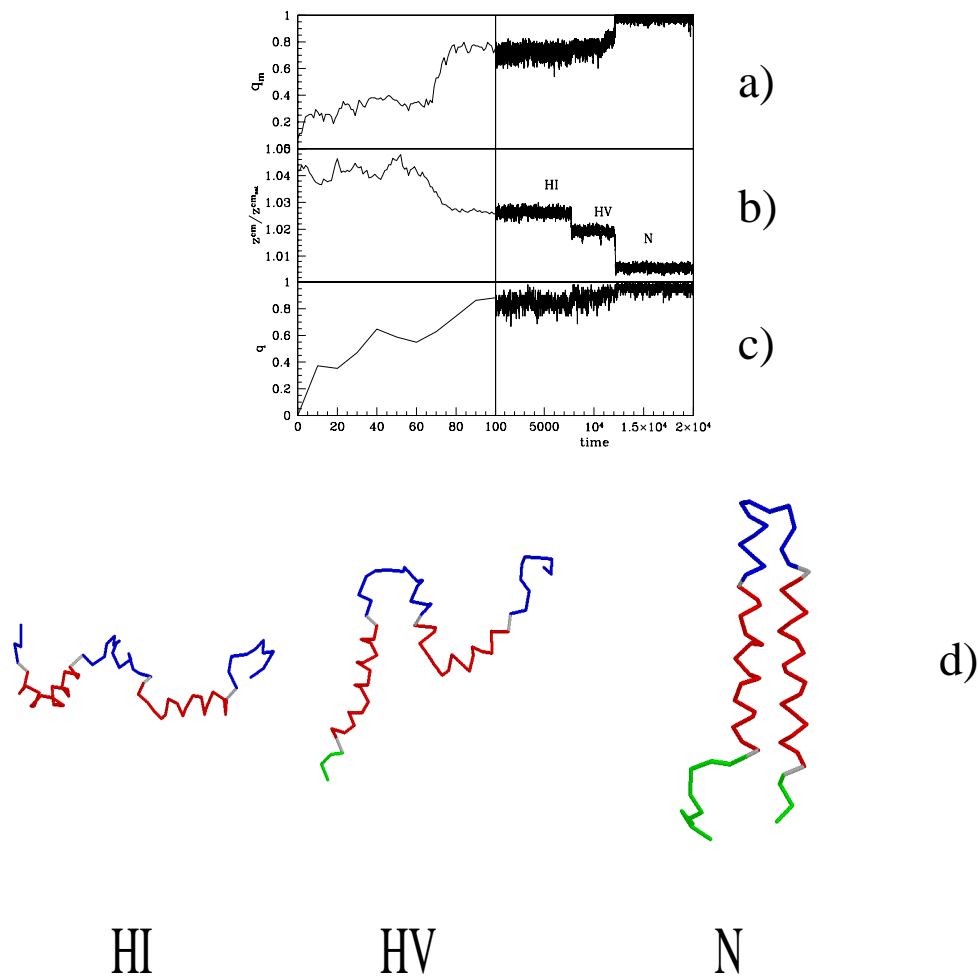}{0.5}
\caption{Typical time dependence of different parameters as a function
of the Monte-Carlo steps for the pathway $U \rightarrow HI \rightarrow
HV \rightarrow N$. Fraction of native contacts inside the membrane
({\bf a}), normalized z-coordinate of the center of mass of the
protein (with respect to that of the native state conformation) ({\bf
b}) and overall fraction of native helical contacts ({\bf c}). ({\bf
d}) Protein conformations at different times during the folding. The
colours red, green and blue have the same significance as in Figure 1a
with the grey bonds being ones crossing the membrane.}
\label{fig:mpfig3}
\end{figure}

In all the pathways, the system goes from the unfolded state, $U$ to
state $ HI$ in which $80\%$ of the secondary structure is formed (see
$q$ in Figure \ref{fig:mpfig3}c) and disposed horizontally along the
interface. The free energy of this state (measured with respect to the
free energy of the fully folded state) is $\sim 2.4$ T$_C$. This state
corresponds to the formation of around $70\ \%$ of the membrane
contacts. The average time $\tau _{HI}$ to reach state $HI$ is of the
order of $500$ Monte Carlo steps (see Figures \ref{fig:mpfig3}; each
Monte-Carlo step corresponds to 50000 attempted local
deformations.). State HI turns out to be an obligatory on-pathway
intermediate of the folding kinetics for non-constitutive MP in
agreement with the general argument mentioned above. Once the protein
reaches state $HI$, it undergoes a relatively slow process of
self-arrangement in order to insert and assemble the secondary
structures across the membrane. This process is the rate-limiting step
of the folding process, since it involves the translocation, through
the lipidic layer, of a substantial number of hydrophilic
residues. Among the possible pathways, starting from $HI$, the most
frequent ($60\%$ of the cases) and the fastest turn out to be $
U\rightarrow HI\rightarrow HV\rightarrow N$. A quantitative
characterization of this dominant pathway is presented in Figures
\ref{fig:mpfig3} (for a single folding process). The intermediate $HV$
is characterized by having one $\alpha $ helix inserted across the
membrane and is reached in an average period corresponding to a
significant fraction of the total folding time (see Figure
\ref{fig:mpfig1}). The free energy in this state is $\sim 0.98$ $T_C.$
The free energy barrier between $HI $ and $HV$ is at $\sim $ $4.31$
$T_C$ (hence, the rate constant of the transition $HI\rightarrow HV$
is proportional to $k_{HI\rightarrow HV}=\exp \left( -\left(
4.31-2.4\right)T_C/T \right) $).

The last part of the folding process corresponds to the insertion of
the second helix and the assembly of the two secondary structures into
the native state structure. This process lasts approximately one third
of the folding time along the pathway $U\rightarrow HI\rightarrow
HV\rightarrow N$. The quasistatic free energy barrier between $HV$
and the folded state is $ \sim $ $1.66$ $T_C$. The rate constant of
the transition $HV\rightarrow N$ is, therefore, proportional to $\exp
\left( -\left( 1.66-0.98\right)T_C/T\right)$. These results are
consistent with the time scales observed in the unconstrained folding
dynamics. At the end, the protein is completely packed, ($q_m$
saturates to 1 (Figure \ref{fig:mpfig3}a) and the helices are
correctly positioned across the membrane (note the second jump in the
$z$ coordinate of the center of mass in Figure \ref{fig:mpfig3}b).

Much slower dynamics can occur when non-obligatory intermediates are
visited by the system. These long lived states ($\{ I \}$ in Figure
\ref{fig:mpfig2}a) involve a distribution of misfolded regions that
trap the system and are characterized by having most of the
inter-helical contacts formed (assembly of the secondary structures)
but with the two ${\alpha }$-helices still incorrectly
positioned. Note, for example, that in states $\{ I \}$ , only
transmembrane contacts and some contacts outside the membrane are
misplaced and they account for only a small fraction of the native
state energy. For this reason, in the states $\{I \} $, the free
energy is $\sim $ $1.44$ $T_C$, only slightly higher than the free
energy of $HV$. The folding can proceed from $\{I \}$ either by
disentangling the two helices and passing through the obligatory
intermediate $HV$, or by the simultaneous translocation through the
membrane of the two helices. These processes, however, entail the
crossing of a big free energy barrier ($\sim $ $5.18$ $T_C$ for the
first process and $6.1$ $T_C$ for the second) and happen with low
probability. Indeed, at sufficiently low temperatures, the loss in
energy of the interhelical contacts is not compensated by the gain in
the configurational entropy due to the uncoupling of the ${\alpha
-}$helices. Thus below the folding temperature, I-states act as
trapping regions for the system and when trapped, the protein spends
most of the time during folding in this state.

In summary, we have shown that a topology based model can lead to a
vivid picture of the folding process. Our approach predicts a folding
process involving multiple pathways with a dominant folding channel.
Further details not captured by the present approach may of course
change the quantitative nature of the results. However, the model,
which captures the bare essentials of a membrane protein, ought to
provide a zeroth order picture of the folding process. Also, as
experimental data becomes available, the results could be benchmarked
with models of this type to determine the nature of the other factors
that matter.


\begin{thebibliography}{0}
\bibitem{cont} The native state structures of monomeric
proteins of length between 50 and 200 show an excellent correlation of
this form, when two non-consecutive amino acids along the sequence are
defined to be in contact when they are within 6.5 \AA\ of each
other.
\bibitem{footnote1} The degree of the overlap may be benchmarked
against the typical overlap of any two compact conformations,
which is about 10-20 \% (consistent with the unfolding simulations of
ref. [\cite{karp}]). A value of $q \approx 20 \%$ thus
corresponds to a ``molten globule'' state.
\end{thebibliography}
\end{document}